\def\year{2021}\relax
\documentclass[letterpaper]{article} 
\usepackage{aaai21}  
\usepackage{times}  
\usepackage{helvet} 
\usepackage{courier}  
\usepackage[hyphens]{url}  
\usepackage{graphicx} 
\usepackage{natbib}  
\usepackage{caption} 
\usepackage{amsfonts}       
\usepackage{nicefrac}       
\usepackage{microtype}      
\usepackage{amsmath}
\usepackage{multicol}
\usepackage[displaymath, mathlines]{lineno}
\usepackage{subfigure}
\usepackage{commath}
\usepackage{amssymb}
\usepackage{nccmath}
\usepackage{float}
\usepackage{bm}
\usepackage{dsfont}
\usepackage{mdframed}
\usepackage{multirow}
\usepackage{tabularx}
\usepackage{tabulary}
\usepackage{amsthm}
\DeclareMathOperator*{\argmax}{argmax} 
\DeclareMathOperator*{\argmin}{argmin} 

\newtheorem{theorem}{Theorem}
\newtheorem{definition}[theorem]{Definition}

\usepackage{tikz}
\usepackage{rotating}
\usepackage{mathtools}
\usepackage{booktabs}
\usepackage{adjustbox}
\usepackage[ruled]{algorithm2e}
\usepackage{tabularx}
\usepackage[vlines]{tabularht}
\usepackage{subfiles} 
\usepackage{lineno}
\usepackage{wrapfig}
\newcommand{\beginsupplement}{%
        \setcounter{table}{0}
        \renewcommand{\thetable}{S\arabic{table}}%
        \setcounter{figure}{0}
        \renewcommand{\thefigure}{S\arabic{figure}}%
     }
     
\iftrue
\usepackage{lastpage}
\usepackage{fancyhdr}
\fi

\title{Low-Rank Reorganization via Proportional Hazards Non-negative Matrix Factorization Unveils Survival Associated Gene Clusters}

\author{
    Zhi Huang,\textsuperscript{\rm 1,2} Paul Salama,\textsuperscript{\rm 3} Wei Shao,\textsuperscript{\rm 2} Jie Zhang, \textsuperscript{\rm 2} Kun Huang\textsuperscript{\rm 2,\thanks{To whom the correspondence should be addressed.}}\\
}
\affiliations{
    \small
    \textsuperscript{\rm 1}School of Electrical and Computer Engineering, Purdue University, West Lafayette, IN, 47907, USA\\
    \textsuperscript{\rm 2} Department of Medicine, Indiana University School of Medicine, Indianapolis, IN, 46202, USA \\
    \textsuperscript{\rm 3} School of Electrical and Computer Engineering, Indiana University - Purdue University Indianapolis, Indianapolis, IN, 46202, USA \\
    huang898@purdue.edu,
    psalama@iupui.edu,
    \{shaowei, jizhan, kunhuang\}@iu.edu
}

\begin{document}

\maketitle

\begin{abstract}
One of the central goals in precision health is the understanding and interpretation of high-dimensional biological data to identify genes and markers associated with disease initiation, development, and outcomes. Though significant effort has been committed to harness gene expression data for multiple analyses while accounting for time-to-event modeling by including survival times, many traditional analyses have focused separately on non-negative matrix factorization (NMF) of the gene expression data matrix and survival regression with Cox proportional hazards model. In this work, Cox proportional hazards regression is integrated with NMF by imposing survival constraints. This is accomplished by jointly optimizing the Frobenius norm and partial log likelihood for events such as death or relapse. Simulation results on synthetic data demonstrated the superiority of the proposed method, when compared to other algorithms, in finding survival associated gene clusters. In addition, using human cancer gene expression data, the proposed technique can unravel critical clusters of cancer genes. The discovered gene clusters reflect rich biological implications and can help identify survival-related biomarkers. Towards the goal of precision health and cancer treatments, the proposed algorithm can help understand and interpret high-dimensional heterogeneous genomics data with accurate identification of survival-associated gene clusters.
\end{abstract}

\section{Introduction}

\noindent Identifying genes and biomarkers associated with patient survival is of great interest to cancer researchers, which has been facilitated by the advances in biotechnology such as next generation sequencing (NGS) as well as the availability of genomics, transcriptomics, proteomics, and other omics data, constituted an indispensable multi-omics source of biological information.

Survival analysis, also known as ``time-to-event'' analysis, aims to model patient lifespan and estimate the time to an event of interest (\textit{e.g.}, death) given the observed data \cite{zahid_2019_survival_analysis}. Survival analysis considers the lifetimes even when the subjects are not experiencing the event of interest, since not all subjects of a given population will experience the event of interest during a study. Cases in which survival times exceeded study duration are labelled as ``censored'' \cite{zahid_2019_survival_analysis}. Though regressing gene expression data via Cox proportional hazards model \cite{cox1972regression} is one of the standard ways to identify survival associated genes, it does not aggregate high-dimensional features nor account for the clustering property to identify gene clusters or pathways. Among the numerous techniques used to discover feature contribution to a classification problem, non-negative matrix factorization (NMF) have demonstrated the dual capability of dimensionality reduction and clustering in latent dimension \cite{ding2005equivalence} which also reflects biological representation \cite{lai2013survival, mejia2008bionmf} by introducing the non-negative constraint.

Non-negative matrix factorization (NMF), studied since 1999 \cite{lee1999learning}, was initially developed for face recognition \cite{zafeiriou2006exploiting, wang2005non, kotsia2007novel, nikitidis2012subclass}, but has since been applied to biological analyses such as gene clustering \cite{pascual2006bionmf, zhu2017robust, zhu2017detecting, jiang2018flexible, lai2013survival} and provide new insights about complex latent relationships in high-dimensional biological data \cite{mejia2008bionmf}. It decomposes a non-negative matrix $\bm X$ into two low-rank matrices: a basis matrix $\bm W$ representing features, and a coefficient matrix $\bm H$ representing samples, provides a well-established geometric and topological representation of the feature space by visualizing the basis matrix. Different from other matrix factorization methods, the imposed non-negative property on $\bm W$ and $\bm H$ can lead to interpretable results \cite{lee1999learning}.

In this paper, by integrating Cox proportional hazards regression into NMF, we developed the ``CoxNMF'' algorithm, demonstrated its superiority among thirty-six different combinations of simulations as well as two real human cancer datasets. To the best of our knowledge, this is the first work that performed non-negative matrix factorization and clustering driven by a simultaneous survival regression. The experiments conducted on human cancer datasets helped unravel latent gene clusters which reflected rich biological interpretations, achieved the goal of understanding and interpretation of high-dimensional biological data in precision health.

\section{Related Work}
\subsection{Cox Proportional Hazards Regression}

Cox proportional hazards regression \cite{cox1972regression} is a well-established survival model assumed that covariates $\bm{X}$ are multiplicatively related to the hazard (death). Suppose given the survival times $0 < Y_1 < Y_2 < Y_3 < \cdots < Y_N$ (assuming no tied times), the survival event $C_i$ for subject $i$ at time $Y_i$, and the matrix $\bm{X}$ denotes the data to be regressed, then the partial likelihood $\mathcal{L}_i(\beta,\bm{X})$ of the death event to be observed for patient $i$ at time $Y_i$ can be written as
\begin{equation}
    \label{eqn:Cox_likelihood_function}
    \small
    \begin{aligned}
    &\mathcal{L}_i(\beta,\bm{X})
    = \frac{\lambda(Y_i|\bm{X}_{\cdot,i})}{\sum_{j:Y_j\geq Y_i}\lambda(Y_i|\bm{X}_{\cdot,j})} \\
    &= \frac{\lambda_0(Y_i)\text{exp}(\beta^T\bm{X}_{\cdot,i})}{\sum_{j:Y_j\geq Y_i}\lambda_0(Y_i)\text{exp}(\beta^T\bm{X}_{\cdot,j})} = \frac{\text{exp}(\beta^T\bm{X}_{\cdot,i})}{\sum_{j:Y_j\geq Y_i}\text{exp}(\beta^T\bm{X}_{\cdot,j})},
    \end{aligned}
\end{equation}
where $\lambda$ is the hazard function \cite{cox1972regression}, and $\beta$ stand for the coefficients associated to the partial likelihood function. The log partial likelihood $\ell_{\bm{X},\beta}(C,Y)$ is derived based on the partial likelihood
\begin{equation}
    \label{eqn:log_partial_likelihood}
    \small
    \ell_{\bm{X},\beta}(C,Y) = \sum_{i:C_i = 1}\left(\beta^T\bm{X}_{\cdot,i} - \text{log}\left( \sum_{j:Y_j\geq Y_i} \text{exp}(\beta^T\bm{X}_{\cdot,j}) \right)\right).
\end{equation}

By optimizing the partial log likelihood $\ell_{\bm{X},\beta}(C,Y)$, the estimated $\beta$ can describe how survival was affected by different covariates in $\bm X$. Basically, a covariate $\bm X_{p,\cdot}$ can lead to worse survival prognosis if the associated $\beta_p>0$ and vice versa. Meanwhile, the concordance index (C-Index), which assesses the model discrimination power of the ability to correctly provide a reliable ranking of the survival times based on the individual risk scores, can be computed as $\text{C-Index} = \frac{\sum_{i,j}\mathds{1}_{(Y_j<Y_i)}\cdot\mathds{1}_{(r_j>r_i)}\cdot C_j}{\sum_{i,j}\mathds{1}_{(Y_j<Y_i)}\cdot C_j}$,
where $r_j$ is the risk score of a subject $j$. $\mathds{1}_{(Y_j<Y_i)}$ is the indicator function: $\mathds{1}_{(Y_j<Y_i)} = 
\begin{cases}
  1 & \text{if } Y_j<Y_i \\
  0 & \text{otherwise}
\end{cases}$.
In computational biology, survival analysis is now one of the most useful analysis to predict patient hazards according to their genomics and transcriptomics data $\bm X$. Such studies can help understand important diagnosis features, signature genes, biomarkers \cite{shao2019diagnosis,shao2020multi,zhang2013network}, and the patients prognosis among various cancers \cite{huang2020deep, ching2018cox}. Survival analysis often conducted after some upfront feature engineering processes such as NMF \cite{jiang2018flexible, jia2015gene} to study biomarkers in clusters.

\subsection{Related Work with NMF}
Applying NMF in biological studies such as unveiling gene interactions and clusters have been exploited since last decade. \cite{liu2008reducing} compared PCA with NMF to reduce the dimension of microarray data and showed the superiority of NMF. \cite{zheng2009tumor} used NMF technique to identify tumor types. \cite{wang2013non} and \cite{gao2005improving} performed cancer clustering with NMF algorithms. Using NMF to perform gene expression clustering can be found in \cite{liu2008alpha, chen2014gene, liu2015hessian, carmona2006biclustering, wang2006ls, pascual2006bionmf, zhu2017robust}. For example, \cite{wang2006ls} proposed LS-NMF to link functionally related genes. More recently, \cite{zhu2017detecting} suggested that NMF is well-suited to analyze heterogeneous single-cell RNA sequencing data. \cite{jiang2018flexible} used NMF to unravel disease-related genes. \cite{jia2015gene} developed discriminant NMF to rank genes. \cite{lai2013survival} performed survival prediction after NMF-based pre-selection of genes. 

Though numerous evidences showed that NMF has an inherent clustering property \cite{ding2005equivalence}, fully utilizing the survival data along with the given gene expression matrix and effectively integrating hazards provided by survival information in its update rule has not been systematically studied. To address this gap, we aim to find an ideal solution to the NMF along with the associated survival data simultaneously. The derived solution successfully demonstrated the power of retrieving survival associated clusters in both synthetic data and human cancer expressions. As a success endeavor towards the central goal of precision health and cancer treatments, the proposed algorithm CoxNMF can help understand and interpret high-dimensional biological data, as well as unveiling critical gene clusters that associated with the survival.

\section{Methodology}

In this section, the algorithm ``CoxNMF'' was proposed and elaborated including the objective function, updating rule, and time \& space complexity analyses. We also carried out the settings for model comparison, simulation with thirty-six different combinations of time to event synthetic data, and evaluation metrics. The analysis for real human cancer data was performed after the simulation studies, aimed to discover the clusters of genes which may play important roles to survival.

\subsection{Variables, Inputs, and Outputs}

It is assumed that the input gene expression data $\bm X$ with shape $P$ by $N$ to be non-negative, real-valued 2-dimensional matrix, which may contains several zero-valued entries. The rows indicate $P$ features/genes, and the columns indicate $N$ samples/patients. The output produced by CoxNMF is consisted of three parts: a low-rank $K$ by $N$ coefficient matrix $\hat{\bm H}$ which learned from NMF and Cox proportional hazards regression, a low-rank $P$ by $K$ basis matrix $\hat{\bm W}$, associated with $\hat{\bm H}$ that minimizes the Frobenius norm $\norm{\bm X - \hat{\bm W}\hat{\bm H}}_{F}$, and the simultaneously learned $K$ by $1$ Cox proportional hazards weight $\hat{\beta}$.

\subsection*{Objective Function}

Given the target non-negative matrix $\bm{X}$, two initialized non-negative matrices $\bm{W}^{(0)}$ and $\bm{H}^{(0)}$, the objective function of CoxNMF is

\begin{equation}
    \label{eqn:CoxNMF_objfunc}
    \small
    \begin{aligned}
    \text{Minimize}\; &\norm{\bm{X}-\bm W \bm H}^2_F\\
    &-\alpha \sum_{i:C_i = 1}\Big(\beta^T\bm{H}_i - \text{log}\big( \sum_{j:Y_j\geq Y_i} \text{exp}(\beta^T\bm{H}_j) \big)\Big) \\
    &+ \frac{N}{2}\xi \big( \gamma \norm{\beta}_1 + (1-\gamma) \norm{\beta}^{2}_{2} \big)
    \end{aligned}
\end{equation}
Subject to $\bm X \in \mathcal{R}_{\geq 0}^{P\times N}, \bm W \in \mathcal{R}_{\geq 0}^{P\times K}, \bm H \in \mathcal{R}_{\geq 0}^{K\times N}$. Where
\begin{equation}
    \label{eqn:smooth_l1}
    \small
    \norm{\beta}_1 \overset{\Delta}{=} \frac{1}{\eta} \big[\log\left(1+\exp(-\eta \beta)\right)+\log\left(1+\exp(\eta \beta)\right)\big]
\end{equation}
is the smooth approximation of the $\norm{\beta}_1$ \cite{schmidt2009optimization, lee2001ssvm}, $\alpha > 0$ is the positive weight imposing Cox proportional hazards regression, $\norm{\cdot}_F$ is the Frobenius norm, also known as Euclidean distance \cite{lee2001algorithms}, $C$ stands for the death events, $Y$ stands for survival times. Weight $\frac{N}{2}\xi$ imposes an elastic net penalty for the objective function, $\xi \geq 0$. $\gamma \in [0,1]$ balance the L1 and L2 ratio in the elastic net penalty. For a better smooth approximation of $\norm{\beta}_1$, a higher $\eta$ is recommended. Through the Equation \ref{eqn:smooth_l1}, we can impose the L1 penalty ratio as well as calculating its second order derivatives $\frac{\partial^2 \norm{\beta}_1}{\partial \beta \otimes \partial \beta'}$. The complete list of symbol descriptions was reported in Table \ref{table:symbol_description}.

\subsection*{CoxNMF Update Rule}

Given the target non-negative matrix $\bm{X}$, two initialized non-negative matrices $\bm{W}^{(0)}$ and $\bm{H}^{(0)}$, an initialized parameter $\beta^{(0)}$ for Cox proportional hazards regression, survival times $Y$ and events $C$, and the maximum number of iterations $M$, we proposed the CoxNMF alternately iterative update algorithm

\begin{equation}
    \small
    \label{eqn:CoxNMF_multiplicative_update_rule_W}
    \bm{W}^{(iter+1)}_{i,j} \leftarrow \bm{W}^{(iter)}_{i,j}\odot
\frac{\bm{X}{\bm{H}^{(iter)}_{i,j}}^T}{\bm{W}^{(iter)}_{i,j}\bm{H}_{i,j}^{(iter)}{\bm{H}^{(iter)}_{i,j}}^T},\\
\end{equation}

\begin{equation}
    \small
    \begin{aligned}
        \label{eqn:CoxNMF_multiplicative_update_rule_beta}
        \beta^{(iter+1)} &\leftarrow \beta^{(iter)}-{\mathcal{H}_{\bm H, \beta, \xi}^{(iter)}}^{-1}g^{(iter)}_{\bm H, \beta, \xi},
    \end{aligned}
\end{equation}

\begin{equation}
    \small
    \label{eqn:CoxNMF_multiplicative_update_rule_H}
    \begin{aligned}
    \bm{H}^{(iter+1)}_{i,j} &\leftarrow \left(\bm{H}^{(iter)}_{i,j} + \frac{\alpha}{2} \text{max}\left\{ \bm{0},  \frac{\partial\ell_{\bm{H}^{(iter)},\beta}(C,Y)}{\partial\bm{H}^{(iter)}}\right\} \right) \\
    &\odot
\frac{{\bm{W}^{(iter+1)}_{i,j}}^T\bm{X}}{{\bm{W}^{(iter+1)}_{i,j}}^T\bm{W}^{(iter+1)}_{i,j}\bm{H}^{(iter)}_{i,j}},
    \end{aligned}
\end{equation}
where
\begin{equation}
\label{eqn:coxph_hessian}
\small
\begin{aligned}
    &\mathcal{H}_{\bm H, \beta, \xi} = \frac{\partial^2 
    \left(-\alpha\ell_{\bm{H},\beta}(C,Y)+\xi\norm{\beta}_1\right)}{\partial \beta \otimes
\partial \beta'}\Biggr|_{\beta=\hat{\beta}^{(iter)}} \\
    &= \alpha\sum _{i:C_{i}=1}\Bigg({\frac {\sum _{j:Y_{j}\geq Y_{i}}\exp(\beta^T\bm{H})_{j}\bm{H}_{j}\bm{H}_{j}^T}{\sum _{j:Y_{j}\geq Y_{i}}\exp(\beta^T\bm{H})_{j}}}\\
    &-{\frac {\left[\sum _{j:Y_{j}\geq Y_{i}}\exp(\beta^T\bm{H})_{j}\bm{H}_{j}\right]\left[\sum _{j:Y_{j}\geq Y_{i}}\exp(\beta^T\bm{H})_{j}\bm{H}_{j}^T\right]}{\left[\sum _{j:Y_{j}\geq Y_{i}}\exp(\beta^T\bm{H})_{j}\right]^{2}}}\Bigg)\\
    &+ \xi\text{diag}\left( \frac{\partial^2\norm{\beta}_1}{\partial \beta \otimes
\partial \beta'}\right)
\end{aligned}
\end{equation}
is the Hessian matrix of Equation \ref{eqn:log_partial_likelihood},
\begin{equation}
\small
\label{eqn:coxph_gradient}
\begin{aligned}
    &g_{\bm H, \beta, \xi} = 
    \frac{\partial \left(-\alpha\ell_{\bm{H},\beta}(C,Y)+\xi\norm{\beta}_1
    \right)}{\partial \beta}\Biggr|_{\beta=\hat{\beta}^{(iter)}}\\ &=-\alpha\sum_{i:C_i=1}\left( \bm{H}_i - \frac{\sum_{j:Y_j\geq Y_i}\text{exp}(\beta^T\bm{H})_j\bm{H}_j }{\sum_{j:Y_j\geq Y_i}\text{exp}(\beta^T\bm{H})_j } \right) + \xi \frac{\partial\norm{\beta}_1}{\partial \beta}
\end{aligned}
\end{equation}
is the partial gradient of Equation \ref{eqn:log_partial_likelihood} with respect to the $\beta$, and
\begin{equation}
    \small
    \begin{aligned}
        \label{eqn:CoxNMF_multiplicative_update_rule_H_detail}
        \frac{\partial\ell_{\bm{H},\beta}(C,Y)}{\partial\bm{H}} &= \begin{bmatrix}
        \frac{\partial\ell_{\bm{H},\beta}(C,Y)}{\partial\bm{H}_{1,1}} & \frac{\partial\ell_{\bm{H},\beta}(C,Y)}{\partial\bm{H}_{1,2}} & \cdots & \frac{\partial\ell_{\bm{H},\beta}(C,Y)}{\partial\bm{H}_{1,N}}\\
        \frac{\partial\ell_{\bm{H},\beta}(C,Y)}{\partial\bm{H}_{2,1}} & \frac{\partial\ell_{\bm{H},\beta}(C,Y)}{\partial\bm{H}_{2,2}} & \cdots & \frac{\partial\ell_{\bm{H},\beta}(C,Y)}{\partial\bm{H}_{2,N}}\\
        \vdots & \vdots & \ddots & \vdots\\
        \frac{\partial\ell_{\bm{H},\beta}(C,Y)}{\partial\bm{H}_{K,1}} & \frac{\partial\ell_{\bm{H},\beta}(C,Y)}{\partial\bm{H}_{K,2}} & \cdots & \frac{\partial\ell_{\bm{H},\beta}(C,Y)}{\partial\bm{H}_{K,N}}\\
        \end{bmatrix}\\
        &= \begin{bmatrix}
        \underbrace{\left(C_r\beta - \sum\limits^{N}_{s=r}C_s\frac{\mathds{1}_{(Y_r\geq Y_s)}\beta\text{exp}(\beta^T\bm{H}_r)}{\sum_{j:Y_j\geq Y_s}\text{exp}(\beta^T\bm{H}_j)}\right)}_\text{ $K\times 1$ vector which repeats $N$ times for $r=1,2,\cdots,N$.}
        \end{bmatrix}
    \end{aligned}
\end{equation}
is the partial derivative of Equation \ref{eqn:log_partial_likelihood} with respect to the $\bm H$, $\mathds{1}_{(Y_i\geq Y_s)} = 
\begin{cases}
  1 & \text{if } Y_i\geq Y_s \\
  0 & \text{otherwise}
\end{cases}$ is the indicator function. Newton-Raphson \cite{kelley1999iterative} as maximum partial likelihood estimator (MPLE) \cite{lee2003statistical} was used for updating Equation \ref{eqn:CoxNMF_multiplicative_update_rule_beta}. The $\hat{\bm W}$ and $\hat{\bm H}$ are returned where the algorithm achieved maximum concordance index (C-Index) during the optimization. The derivation of Equation \ref{eqn:CoxNMF_multiplicative_update_rule_H} was further elaborated in the \textbf{Appendix} Definition \ref{def:1}.

In practice, Efron's method \cite{efron1977efficiency} was adopted as the efficient alternatives of Equation \ref{eqn:coxph_hessian} and \ref{eqn:coxph_gradient}. Thus we have
\begin{equation}
\small
\label{eqn:coxph_hessian_Efron}
\begin{aligned}
    \mathcal{H}_{\bm H, \beta, \xi} &\overset{\Delta}{=} \alpha\sum _{j=1}^{N}\sum _{\ell =0}^{m_j-1}\Bigg({\frac {\sum _{i:Y_{i}\geq Y_{j}}(\beta^{T}\bm{H})_{i}\bm{H}_{i}\bm{H}_{i}^T}{\phi _{j,\ell ,m_j}}} \\
    &-\frac{{\frac {\ell }{m_j}}\sum _{i\in \Psi_{j}}(\beta^{T}\bm{H})_{i}\bm{H}_{i}\bm{H}_{i}^T}{\phi _{j,\ell ,m_j}}-{\frac {Z_{j,\ell ,m_j}Z_{j,\ell ,m_j}^T}{\phi _{j,\ell ,m_j}^{2}}}\Bigg)\\
    &+ \xi\text{diag}\left( \frac{\partial^2\norm{\beta}_1}{\partial \beta \otimes
\partial \beta'}\right),
\end{aligned}
\end{equation}
and
\begin{equation}
\small
\begin{aligned}
    \label{eqn:coxph_gradient_Efron}
    &g_{\bm H, \beta, \xi} \overset{\Delta}{=} -\alpha\sum _{j=1}^{N}\Bigg(\sum _{i\in \Psi_{j}}\bm{H}_{i}\\
    &-\sum _{\ell =0}^{m_j-1}{\frac {\sum _{i:Y_{i}\geq Y_{j}}(\beta^{T}\bm{H})_{i}\bm{H}_{i}-{\frac {\ell }{m_j}}\sum _{i\in \Psi_{j}}(\beta^{T}\bm{H})_{i}\bm{H}_{i}}{\sum _{i:Y_{i}\geq Y_{j}}\beta^{T}\bm{H} _{i}-{\frac {\ell }{m_j}}\sum _{i\in \Psi_{j}}\beta^{T}\bm{H} _{i}}}\Bigg)\\
    &+ \xi \frac{\partial\norm{\beta}_1}{\partial \beta},
\end{aligned}
\end{equation}
where 
\begin{equation}
\small
    \phi _{j,\ell ,m_j}=\sum _{i:Y_{i}\geq Y_{j}}(\beta^{T}\bm{H})_{i}-{\frac {\ell }{m_j}}\sum _{i\in \Psi_{j}}(\beta^{T}\bm{H})_{i},
\end{equation}
\begin{equation}
\small
    Z_{j,\ell ,m_j}=\sum _{i:Y_{i}\geq Y_{j}}(\beta^{T}\bm{H})_{i}\bm{H}_{i}-{\frac {\ell }{m_j}}\sum _{i\in \Psi_{j}}(\beta^{T}\bm{H})_{i}\bm{H}_{i},
\end{equation}
$\Psi_j$ denote the set of indices $i$ such that $Y_i = Y_j$ and $C_i = 1$, and $m_j$ is the number of the death count at $Y_j$. Pseudo code is presented below.

\subsubsection{Time and Space Complexities}
\begin{figure*}[h]
\begin{center}
  \includegraphics[width=0.4\linewidth]{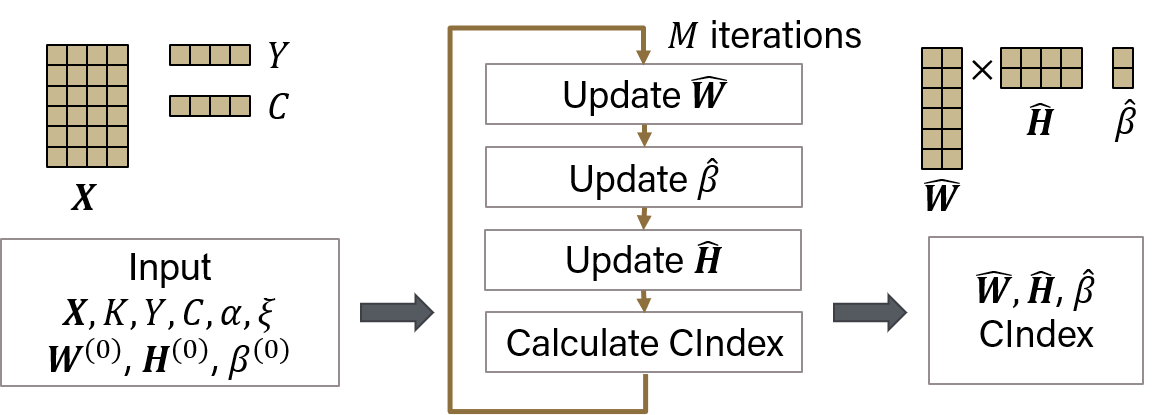}
  \caption{%
    Flowchart of the proposed CoxNMF algorithm.
    }
  \label{fig:flowchart}
\end{center}
\end{figure*}

The time complexity of CoxNMF for one iteration consists of Equation \ref{eqn:CoxNMF_multiplicative_update_rule_W}, \ref{eqn:CoxNMF_multiplicative_update_rule_beta}, and \ref{eqn:CoxNMF_multiplicative_update_rule_H} is $O(PNK+N^2K+K^2\max(P,N))$, the space complexity is $O(PN+N^2)$. Proofs of the time and space complexities were elaborated in the \textbf{Appendix} Theorem \ref{thm:time_complexity} and \ref{thm:space_complexity}. A flowchart visualized CoxNMF updating rule can be found in Figure \ref{fig:flowchart}.

\begin{algorithm}[h]
\small
\SetKwInOut{Input}{Input}
\SetKwInOut{Output}{Output}
\SetKwInOut{Initialization}{Initialization}
\SetKw{Continue}{continue}
\caption{\textsc{CoxNMF}}
\Input{$\bm X$, $K$, $Y$, $C$, $\alpha$, $M$.}
\Output{$\bm W$, $\bm H$, $\beta$, and $CI$.}
\Initialization{Initialize $\bm W^{(0)}$, $\bm H^{(0)}$, $\beta^{(0)}$, empty list $CI$.}

\For{$iter=0:M-1$}{

    $\bm{W}^{(iter+1)} \leftarrow \bm{W}^{(iter)}\odot
\frac{\bm{X}{\bm{H}^{(iter)}}^T}{\bm{W}^{(iter)}\bm{H}^{(iter)}{\bm{H}^{(iter)}}^T}$

    $\beta^{(iter+1)} \leftarrow \beta^{(iter)}-{\mathcal{H}_{\bm H, \beta, \xi}^{(iter)}}^{-1}g^{(iter)}_{\bm H, \beta, \xi}$

    $\bm{H}^{(iter+1)} \leftarrow \left(\bm{H}^{(iter)} + \frac{\alpha}{2} \text{max}\left\{ \bm{0}, \frac{\partial\ell_{\bm{H}^{(iter)},\beta}(C,Y)}{\partial\bm{H}^{(iter)}}\right\} \right) \odot \frac{{\bm{W}^{(iter+1)}}^T\bm{X}}{{\bm{W}^{(iter+1)}}^T\bm{W}^{(iter+1)}\bm{H}^{(iter)}}$

    $\text{CI}[iter+1] = \textsc{ConcordanceIndex}({\beta^{(iter+1)}}^T\bm{H}^{(iter+1)}, Y, C)$
}
$imax = \argmax_{(iter)}{CI}$\\
$\hat{\bm W} = \bm W^{(imax)}$\\
$\hat{\bm H} = \bm H^{(imax)}$\\
$\hat{\beta} = \beta^{(imax)}$\\
$CI = CI[imax]$\\
\Return{$\hat{\bm W}, \hat{\bm H}, \hat{\beta}, CI$}
\end{algorithm}

\subsection{Model Setup and Comparisons}
\subsubsection{Unconstrained Low-Rank Approaches}
Starting from unconstrained matrix factorization approaches, truncated singular value decomposition (SVD) \cite{halko2011finding}, principal component analysis (PCA) \cite{martinsson2011randomized}, sparse PCA \cite{d2005direct}, factor analysis (FA) \cite{everett2013introduction}, and nonnegative double singular value decomposition (NNDSVD) \cite{boutsidis2008svd} were adopted as baseline methods to perform dimensionality reduction upfront, then used the decomposed $\hat{\bm W}$ to cluster genes (features), and the decomposed $\hat{\bm H}$ for survival analysis. All unconstrained methods imposed L1 norm on $\beta$ with weight searching $\xi \in \{0,0.1,1\}, \gamma = 1$. For the sparse PCA, we tuned the sparsity controlling parameter in range $\{0.1,1,10,100,1000\}$ \cite{jenatton2010structured}. We denoted these low-rank approaches as ``unconstrained'' since they do not impose non-negativity in basis and coefficient matrices, and the Cox regression is performed afterwards.

\subsubsection{Non-negative Matrix Factorization Approaches}
CoxNMF was further compared with two vanilla NMF with Coordinate Descent (CD) solver and Mutiplicative Update (MU) solver \cite{lee1999learning}. In addition, CoxNMF was compared to supervised non-negative matrix factorization (SNMF) \cite{chao2018supervised} which imposed linear regression into the NMF optimization process. In SNMF, survival times $t$ are regressed. Unlike CoxNMF which simultaneously minimize the Frobenius norm and maximize the partial log likelihood, methods with NMF (CD), NMF (MU), and SNMF performed Cox proportional hazards regression on $\hat{\bm H}$ afterwards. All NMF-based methods imposed L1 norm on $\beta$ with weight searching $\xi \in \{0,0.1,1\}, \gamma = 1$.  We tuned $\alpha=\frac{v}{P}, \beta=\frac{v}{K}, \gamma=\frac{v}{N\times K}, v \in \{0.001,0.01,0.1\}$ for SNMF according to \cite{chao2018supervised}, $\alpha \in \{0.1,1,2,5,10,20,50,100,200,500,1000\}$ for CoxNMF, and used NMF(CD), NMF (MU), NNDSVD, or random as initializer for $\bm{W}^{(0)}$ and $\bm{H}^{(0)}$ in CoxNMF among all experiments, where $P, N, K$ stands for number of features, patients, and low-rank dimensions, respectively. In this paper, all algorithms including CoxNMF set the maximum number of iterations $M=500$.

\subsection{Simulation Settings with Synthetic Data}

In the simulation, we considered the dataset contains $N=100$ patients. The survival time $t_i$ of patient $i$ were constructed in ascending order follows exponential distribution $\lambda e^{-\lambda t}$ with $\lambda=100$, $i \in \{1,2,\cdots,N\}$ according to \cite{bender2005generating}. All patients will have a complete event ($C_i = 1$) in this ideal situation.

With $K$ number of latent gene clusters each consists of $50$ genes (ground truth), we have the ground truth basis matrix $\bm W$ in block diagonal with dimension $P$ by $K$, each block was non-negative and assumed to follow \textit{i.i.d.} exponential distribution $\phi e^{-\phi w}$ with $\phi=1$ \cite{cox1972regression}. In this case, $P = 50\cdot K$. We denote $\bm{W}_{[k]}$ as the $k^{\text{th}}$ block consists of 50 genes in $k^{th}$ cluster.

The ground truth coefficient matrix $\bm H$ with dimension $K$ by $N$, each row $k \in \{1,\cdots,K\}$ of $\bm{H}_{k,\cdot}$ follows \textit{i.i.d.} uniform distribution $\mathcal{U}(0,1)$ except certain rows carried survival information. we supposed the $\bm{H}_{i,\cdot}$ for $i \in \{1,2,\cdots, \tau_1\}$ are associated with better prognosis (concord with survival time), and $\bm{H}_{j,\cdot}$ for $j \in \{K-\tau_2+1,\cdots, K\}$ are associated with worse prognosis (discord with survival time), $\tau_1 + \tau_2 < K$.

\subsubsection{Univariate Underlying Features}
In this setting, the first/last rows were artificially replaced ($\tau_1=\tau_2=1$). Specifically, the value of first row $\bm{H}_{k=1,i}$ was constructed from 0.5 to 0.698 with step size = $0.02$ (C-Index = 0). The value of last row $\bm{H}_{k=K,i}$ was constructed from 0.698 to 0.5 with step size = $-0.02$ (C-Index = 1). This simulation setting hypothesized that only first/last row carried survival information which concord/discord with survival. In this case, our goal is to unravel the gene clusters $\bm W_{[1]}$ and $\bm W_{[K]}$ associated with better/worse survival prognosis.

\subsubsection{Multivariate Underlying Features}
In reality, the underlying group of features may dependent on each other. To simulate this situation, we set $\tau_1=\tau_2=2$. Initially all rows of the $\bm{H}$ follow \textit{i.i.d.} uniform distribution $\mathcal{U}(0,1)$. Then the second row $\bm{H}_{k=2,i}$ was replaced by the array from 0.5 to 0.698 with step size = $0.02$, further minus $\frac{1}{10}\bm{H}_{k=1,i}$. Meanwhile, the $(K-1)^{\text{th}}$ row $\bm{H}_{k=K-1,i}$ was replaced by the array from 0.698 to 0.5 with step size = $-0.02$, further minus $\frac{1}{10}\bm{H}_{k=K,i}$. Thus the first and second rows are dependent with each other and both lead to the better prognosis, also vise versa for the $K^{\text{th}}$ and $(K-1)^{\text{th}}$ rows.

The ground truth data matrix $\bm X$ was then constructed as $\bm{X} = \bm{W}\bm{H} + \bm{E}$. The matrix $\bm E$ suggests an artificial noise introduced into the system followed exponential distribution $\varepsilon e^{-\varepsilon w}$ with $\varepsilon\in \{0,0.05,0.1\}$. As in reality $K$ remains unknown, the first step is required to determine the optimal $\hat{K}$ according to the silhouette score (introduced in \textbf{Evaluation Metrics} section). We performed experiments with all combinations of $K \in \{7,8,9,10,11,12\}$ and searching $\hat{K} = K\pm\{0,1,2\}$. All experiments ran 5 times each with different random seeds for the initialization and optimizations. From Table \ref{table:supplement_univariate} and Table \ref{table:supplement_multivariate} we observed almost all algorithms can found the optimal $\hat{K}=K$ based on their highest silhouette score especially the proposed CoxNMF. This step helped us to determine number of the latent dimension $K$ in the simulation study and in human cancer datasets as well.

\begin{figure*}[t]
\begin{center}
  \includegraphics[width=\linewidth]{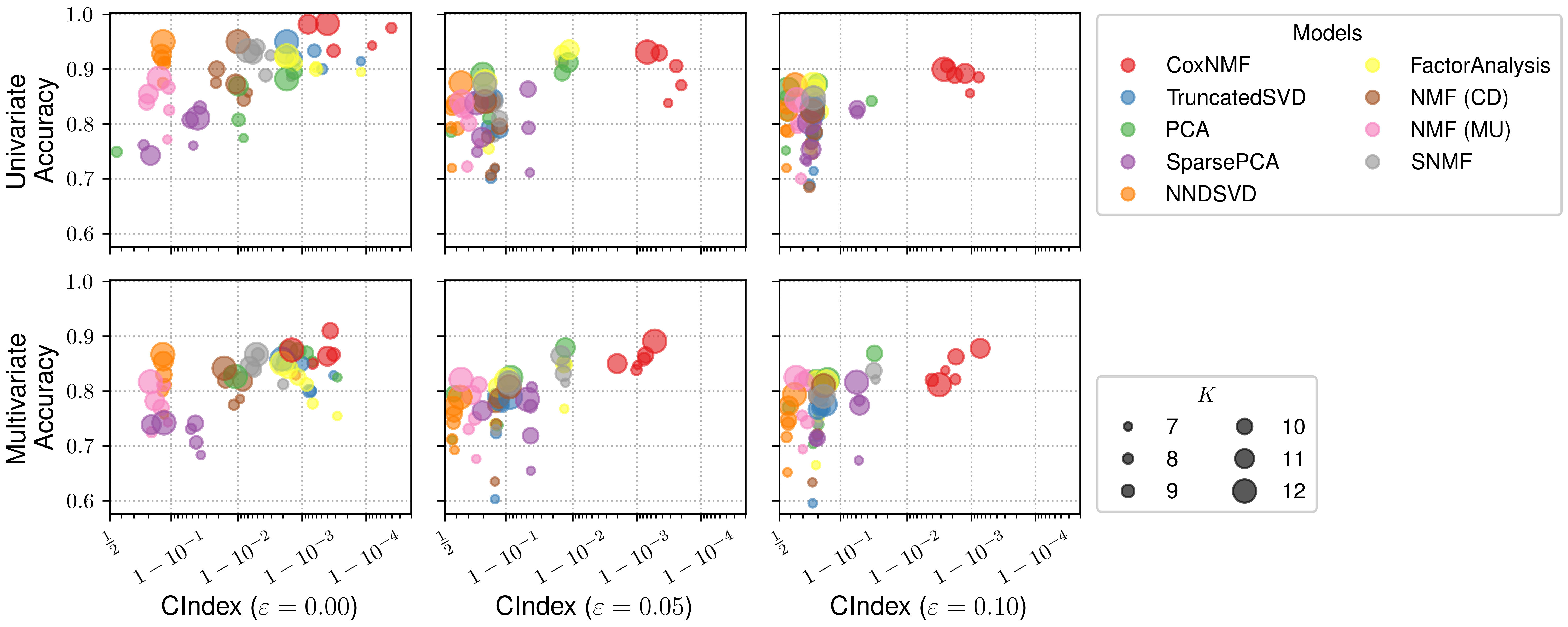}
  \caption{C-Index and accuracy among five unconstrained low-rank approaches and four NMF-based approaches across $K\in\{7,8,9,10,11,12\}$, and three different levels of artificial noise $\bm E$ for $\varepsilon\in\{0,0.05,0.10\}$ in both univariate and multivariate simulations. Mean values from 5 random seeds results were used for presenting this figure. Figure best viewed in color. The performance of C-Index v.s. relative error were also demonstrated in Figure \ref{fig:simulation_CI_vs_rerror}.}
  \label{fig:simulation_CI_vs_acc}
\end{center}
\end{figure*}

\subsection{Evaluation Metrics}

After row normalization on the resulting basis matrix $\bar{\bm W}_p = \hat{\bm W}_p/\norm{\hat{\bm W}_p} $ for each row $p$, we applied the Cox proportional hazards regression parameter weighted on the normalized basis matrix: $\Tilde{\bm W}_p = \hat{\beta}^T\odot\bar{\bm W}_p$ (\textit{i.e.}, the $\hat{\beta}$ parameter will multiply each row $p$ of $\bar{\bm W}$ element-wisely), and sort the columns of $\Tilde{\bm W}$ and rows of $\hat{\bm H}$ according to $\hat{\beta}$ in ascending order, respectively. Note that $\hat{\beta}_k < 0$ suggests a reduction in hazard at $k^{\text{th}}$ latent space, and the elements of $\Tilde{\bm W}$ can be negative. In this case, a negative value in $\Tilde{\bm W}$ suggests an association with better prognosis and vise versa.

\subsubsection{Relative Error of Frobenius Norm}
We introduced the Frobenius norm $\norm{\cdot}_F$ at Equation \ref{eqn:CoxNMF_objfunc}, which is the metric to evaluate the performance of NMF algorithms. Furthermore, we used the relative error \cite{kawaguchi2018non}, which is the ratio of $\norm{\bm X-\hat{\bm W}\hat{\bm H}}_F$ to $\norm{\bm X}_F$ in percentage, to consider whether a low-rank decomposition is adequately learned.

\subsubsection{Silhouette Score for Determining Optimal Number of Latent Dimension $K$}
The silhouette score, or mean silhouette coefficient, measures the consistency within clusters of data. The score describes how well each element has been classified \cite{rousseeuw1987silhouettes}. In this study, Euclidean distance was adopted as distance metric. The hierarchical agglomerative clustering measured in Euclidean distance with Ward linkage \cite{ward1963hierarchical} was performed on $\bar{\bm W}$ to determine the optimal $\hat{K}$ according to the highest silhouette score and gene clusters $L_j, j\in \{1,2,\cdots,\hat{K}\}$. Note that all models can only produce a low-rank basis matrix $\hat{\bm W}$, the hierarchical clustering was to further generate gene clusters given the normalized $\bar{\bm W}$.

\subsubsection{Quantitative Measurements of Optimization Results and Label Accuracy}
To tune the hyper-parameters as well as evaluating the performance of optimization results along with survival information, the concordance index (C-Index) was adopted and defined in the \textbf{Related Work} section. It is a generalization of the area under the ROC curve (AUC) which introduces the censorship information. Similar to the AUC, $\text{C-Index}=1$ corresponds to the best model prediction, and $\text{C-Index}=0.5$ represents a random prediction.

In simulation studies, to quantify whether models can identify the survival-associated gene clusters correctly, we adopted four measurements to evaluate the results, namely, accuracy, F-1 score, precision, and recall. The F-1 score is formulated as $\text{F-1} = 2(\text{precision} \times \text{recall}) / (\text{precision} + \text{recall})$.

To find the survival associated gene clusters, we focused on the $\tau_1$ smallest $\hat{\beta}_{1},\cdots,\hat{\beta}_{\tau_1}$ and $\tau_2$ largest $\hat{\beta}_{K-\tau_2+1}, \cdots, \hat{\beta}_{K}$ associated with $\Tilde{\bm W}$. The true labels $y_{-}$ (or $y_{+}$) for better (or worse) prognosis genes were indicated as 1 at where the genes reside at ${\bm W}_{[1,\cdots,\tau_1]}$ (or at ${\bm W}_{[K-\tau_2+1,\cdots,K]}$). The estimated labels $\hat{y}_{-} = \sum_{i=1}^{\tau_1} \argmin_{L_j} \frac{\sum_j{\Tilde{\bm W}_{L_j=i,i}}}{\sum_{j:
L_j=i}1}$ and $\hat{y}_{+} = \sum_{i=K-\tau_2+1}^{K} \argmax_{L_j} \frac{\sum_j{\Tilde{\bm W}_{L_j=i,i}}}{\sum_{j:
L_j=i}1}$ are determined by the highest mean absolute value on the associated low-rank dimensions (Operation $\argmin_{L_j}$ and $\argmax_{L_j}$ return binary label arrays where genes at $L_j$ were indicated as 1). We concatenated $y_{-}$ and $y_{+}$ as $y$, and concatenated $\hat{y}_{-}$ and $\hat{y}_{+}$ as $\hat{y}$. Then we compared our true labels $y$ and the estimated labels $\hat{y}$ via four metrics: accuracy, F-1 score, precision, and recall to quantify the performances of finding survival-associated gene clusters. Note that these four metrics are only valid for simulation study due to the absence of the ground truth labels in reality. C-Index was determined during the model learning, while accuracy, F-1 score, precision, and recall were unseen until the model with optimal C-Index was determined.
\section{Results}

Our method was first validated on totally thirty-six different combinations of simulation study with univariate and multivariate underlying features setup, six different sizes of synthetic data ($K \in \{7,8,9,10,11,12\}$), further perturbed by three different artificially induced noises ($\varepsilon\in \{0,0.05,0.10\}$). We then applied the proposed method on two real human cancer datasets, Kidney Renal Clear Cell Carcinoma (KIRC), and Lung Adenocarcinoma (LUAD).

\begin{table*}[ht!]
    \centering
    \small
    \begin{tabular}{llllllll}
    \toprule
             & $K$  &            7 &            8 &            9 &           10 &           11 &           12 \\
    Metrics & Model &              &              &              &              &              &              \\
    \midrule
    CIndex & TruncatedSVD &  \textbf{0.9999$\pm$0.00} &  0.9995$\pm$0.00 &  0.9994$\pm$0.00 &  0.9987$\pm$0.00 &  0.9982$\pm$0.00 &  0.9983$\pm$0.00 \\
             & PCA &  0.9918$\pm$0.00 &  0.5590$\pm$0.04 &  0.9902$\pm$0.01 &  0.9987$\pm$0.00 &  0.9901$\pm$0.00 &  0.9983$\pm$0.00 \\
             & SparsePCA &  0.9522$\pm$0.01 &  0.7700$\pm$0.12 &  0.9620$\pm$0.01 &  0.9467$\pm$0.02 &  0.8106$\pm$0.16 &  0.9588$\pm$0.01 \\
             & NNDSVD &  0.8819$\pm$0.02 &  0.8697$\pm$0.01 &  0.8745$\pm$0.02 &  0.8655$\pm$0.01 &  0.8643$\pm$0.01 &  0.8700$\pm$0.02 \\
             & FactorAnalysis &  0.9999$\pm$0.00 &  0.9994$\pm$0.00 &  0.9994$\pm$0.00 &  0.9987$\pm$0.00 &  0.9982$\pm$0.00 &  0.9983$\pm$0.00 \\
             & NMF (CD) &  0.9931$\pm$0.01 &  0.9783$\pm$0.03 &  0.9919$\pm$0.01 &  0.9789$\pm$0.01 &  0.9892$\pm$0.01 &  0.9901$\pm$0.01 \\
             & NMF (MU) &  0.8869$\pm$0.10 &  0.8932$\pm$0.05 &  0.8903$\pm$0.10 &  0.7887$\pm$0.08 &  0.7979$\pm$0.16 &  0.8533$\pm$0.12 \\
             & SNMF &  0.9985$\pm$0.00 &  0.9969$\pm$0.00 &  0.9963$\pm$0.00 &  0.9949$\pm$0.00 &  0.9941$\pm$0.00 &  0.9930$\pm$0.00 \\
             & CoxNMF &  \textbf{0.9999$\pm$0.00} &  \textbf{1.0000$\pm$0.00} &  \textbf{0.9997$\pm$0.00} &  \textbf{1.0000$\pm$0.00} &  \textbf{0.9992$\pm$0.00} &  \textbf{0.9996$\pm$0.00} \\
    \midrule
    Accuracy & TruncatedSVD &  0.9143$\pm$0.08 &  0.9000$\pm$0.06 &  \textbf{0.9333$\pm$0.06} &  0.9200$\pm$0.04 &  0.9273$\pm$0.04 &  0.9500$\pm$0.05 \\
             & PCA &  0.7740$\pm$0.05 &  0.7490$\pm$0.11 &  0.8073$\pm$0.02 &  0.8962$\pm$0.03 &  0.8689$\pm$0.02 &  0.8822$\pm$0.09 \\
             & SparsePCA &  0.7603$\pm$0.05 &  0.7612$\pm$0.07 &  0.8298$\pm$0.01 &  0.8072$\pm$0.06 &  0.7425$\pm$0.06 &  0.8108$\pm$0.09 \\
             & NNDSVD &  0.9143$\pm$0.08 &  0.8750$\pm$0.00 &  0.9111$\pm$0.05 &  0.9200$\pm$0.04 &  0.9273$\pm$0.04 &  0.9500$\pm$0.05 \\
             & FactorAnalysis &  0.8946$\pm$0.04 &  0.9037$\pm$0.06 &  0.8991$\pm$0.07 &  0.9110$\pm$0.07 &  0.9202$\pm$0.04 &  0.9240$\pm$0.04 \\
             & NMF (CD) &  0.8571$\pm$0.00 &  0.8750$\pm$0.09 &  0.8444$\pm$0.06 &  0.9000$\pm$0.00 &  0.8727$\pm$0.05 &  0.9500$\pm$0.05 \\
             & NMF (MU) &  0.7714$\pm$0.08 &  0.8250$\pm$0.07 &  0.8667$\pm$0.09 &  0.8400$\pm$0.05 &  0.8545$\pm$0.05 &  0.8833$\pm$0.05 \\
             & SNMF &  0.8857$\pm$0.06 &  0.9250$\pm$0.07 &  0.8889$\pm$0.00 &  \textbf{0.9400$\pm$0.05} &  0.9273$\pm$0.04 &  0.9333$\pm$0.04 \\
             & CoxNMF &  \textbf{0.9429$\pm$0.08} &  \textbf{0.9750$\pm$0.06} &  \textbf{0.9333$\pm$0.06} &  \textbf{0.9400$\pm$0.05} &  \textbf{0.9818$\pm$0.04} &  \textbf{0.9833$\pm$0.04} \\
    \bottomrule
    \end{tabular}
    
    \caption{Simulation results with univariate underlying features setup for $K\in\{7,8,9,10,11,12\}$, $\varepsilon = 0$. $\hat{K}$ was searched around $K\pm\{0,1,2\}$ and was determined by highest silhouette score. Experiments repeat 5 times each with random seed $\in\{1,2,3,4,5\}$. Mean values $\pm$ standard deviations were reported, best performed mean values among models were highlighted in bold font.}
    \label{table:main_univariate_X_noise=0}
\end{table*}

\subsection*{Simulation Results}

\begin{figure*}[h]
\begin{center}
  \includegraphics[width=0.9\linewidth]{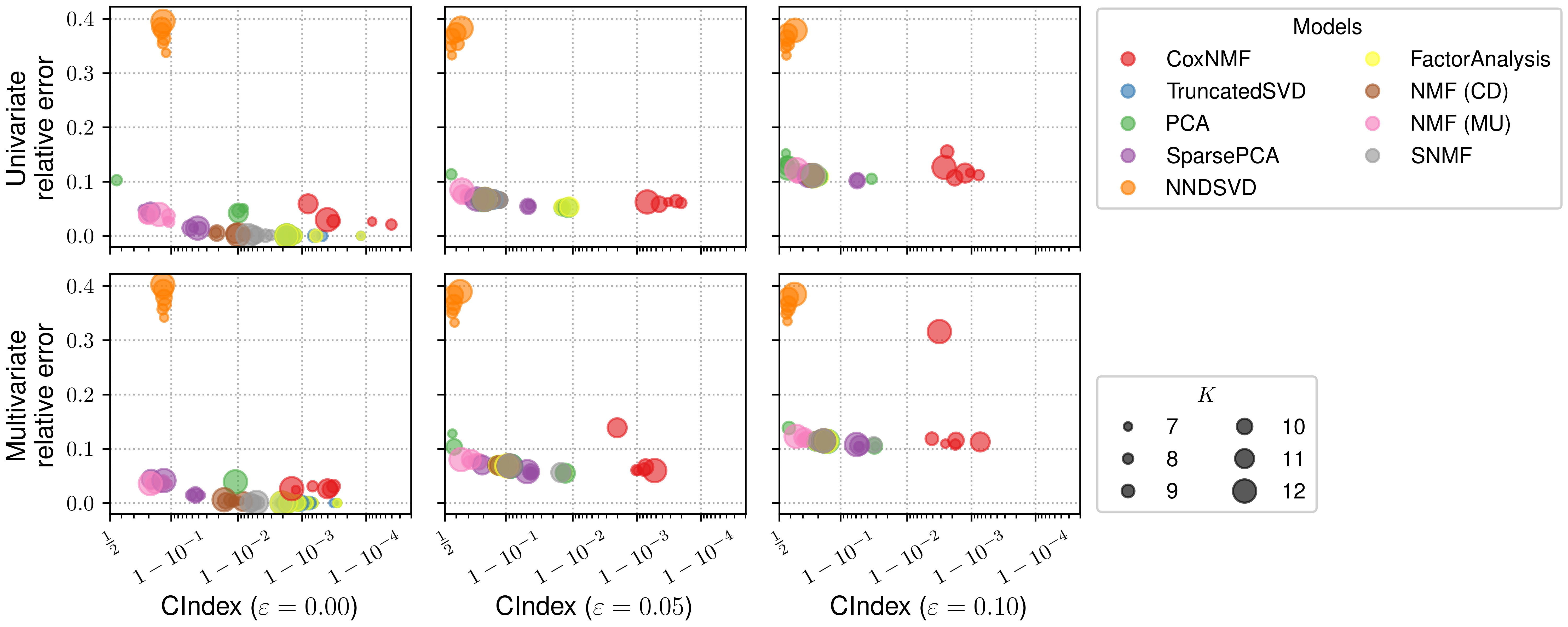}
  \caption{C-Index and relative error among five unconstrained low-rank approaches and four NMF-based approaches across $K\in\{7,8,9,10,11,12\}$, three different levels of artificial noise $\bm E$ for $\varepsilon\in\{0,0.05,0.10\}$ in both univariate and multivariate simulations. Mean values from 5 random seeds results were used for presenting this figure. Figure best viewed in color.}
  \label{fig:simulation_CI_vs_rerror}
\end{center}
\end{figure*}

\begin{figure*}[h]
\begin{center}
  \includegraphics[width=0.9\linewidth]{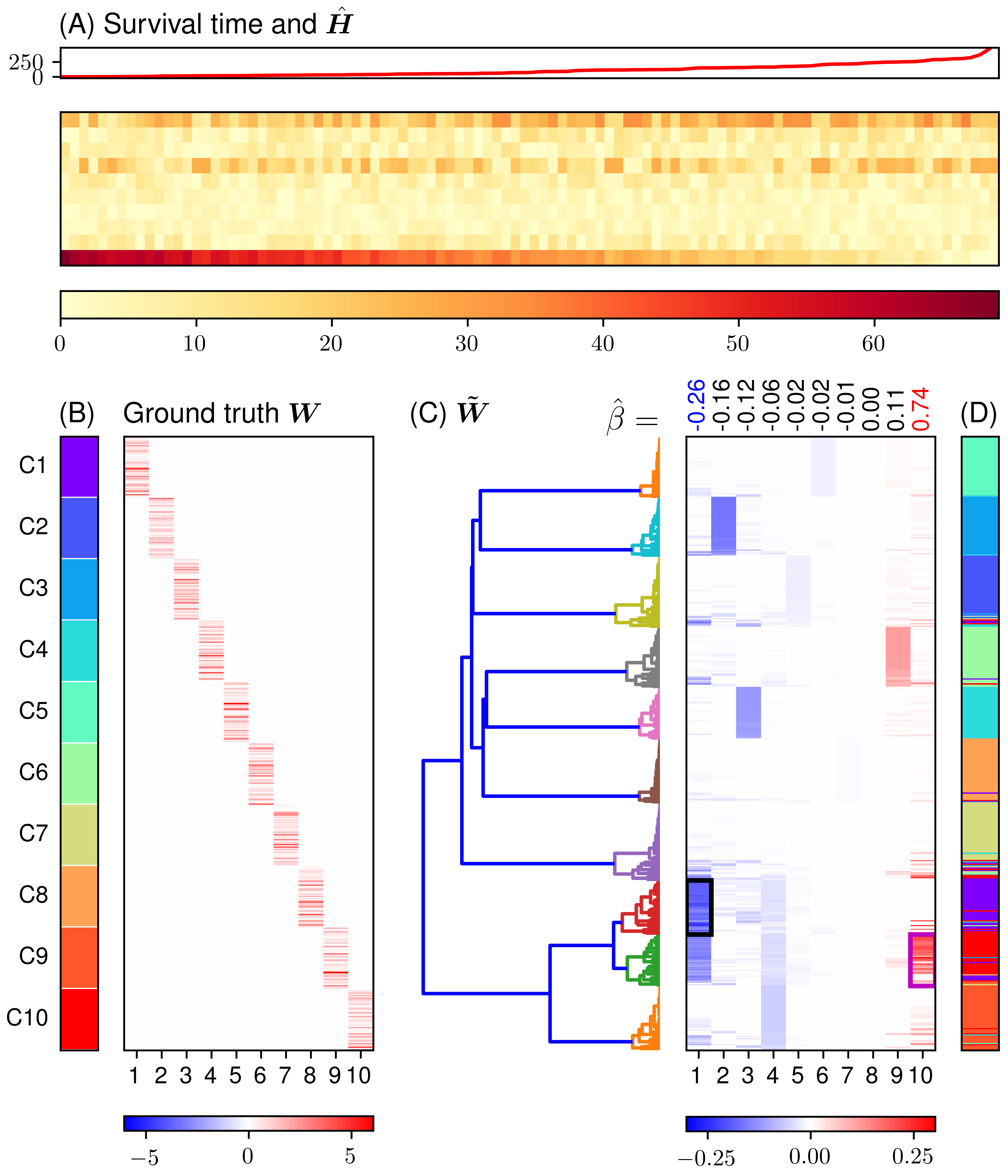}
  \caption{%
    An univariate underlying features simulation result with CoxNMF model ($K=\hat{K}=10$, $\varepsilon = 0.10$). (A) Survival time and $\hat{\bm H}$. (B) ground truth $\bm W$, note that $\bm{W}_{[1]}$ associated with better prognosis (longer survival time), $\bm{W}_{[K]}$ associated with worse prognosis. (C) $\Tilde{\bm W}$ and hierarchical agglomerative clustering results (highlighted by most distinct colors, but did not relate with colors in (B) and (D)) with $\hat{K}$ number of clusters. Columns of $\Tilde{\bm W}$ and rows of $\hat{\bm H}$ were sorted in ascending order of $\hat{\beta}$. Cluster with highest mean value on the smallest $\hat{\beta}_{1}$ and cluster with highest mean value on the largest $\hat{\beta}_{K}$ were highlighted with black rectangle and \textcolor{magenta}{magenta} rectangle. (D) Ground truth labels in plot B with row permutation according to the hierarchical clustering result. In this figure, C-Index $=0.9996$, accuracy $=0.9520$, F-1 score $=0.8577$, precision $=0.8842$, recall $=0.8356$, relative error $=10.6478\%$, and running time $=3.9074$ seconds.}
  \label{fig:simulation_2}
\end{center}
\end{figure*}

\begin{figure*}[h]
\begin{center}
  \includegraphics[width=0.9\linewidth]{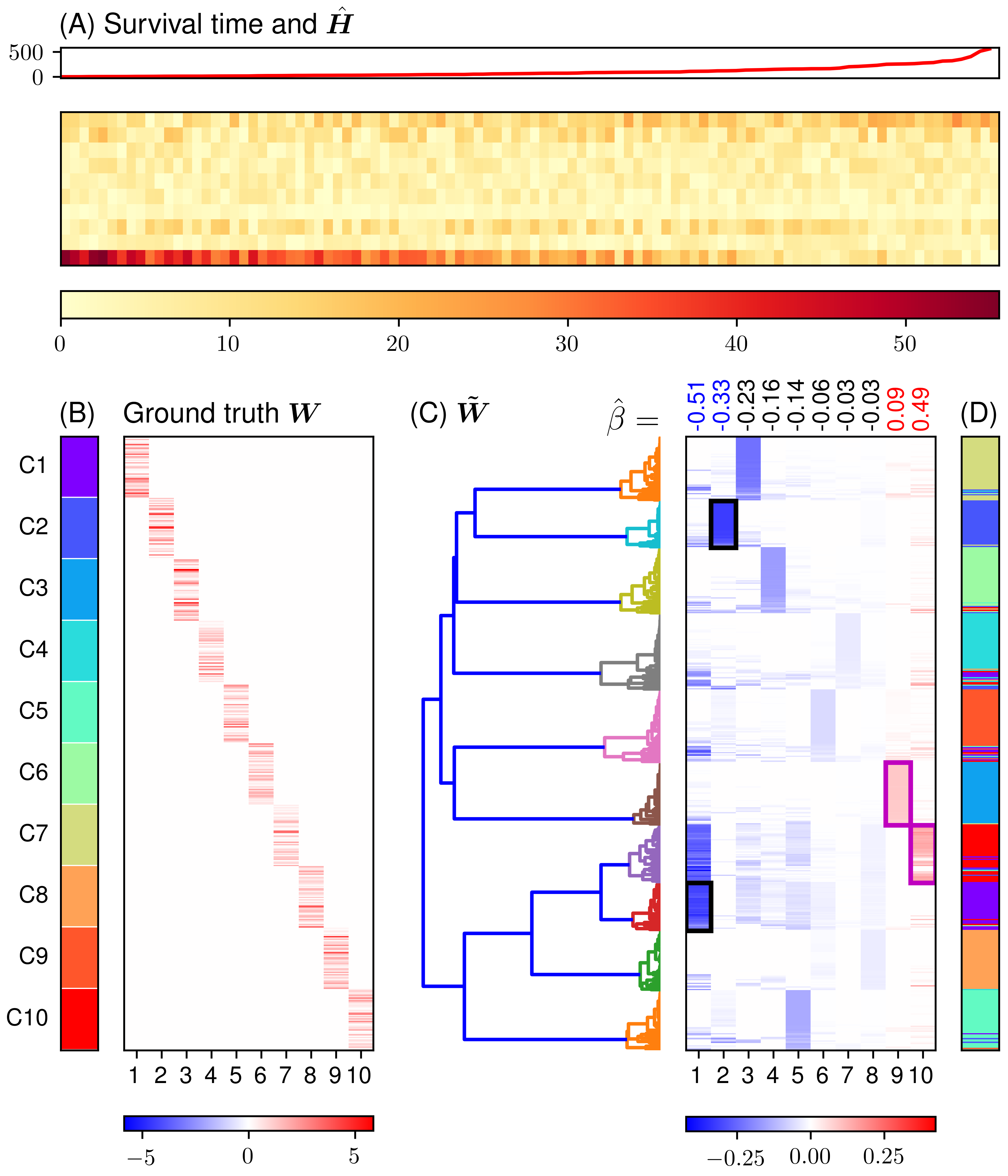}
  \caption{%
    A multivariate underlying features simulation result with CoxNMF model ($K=\hat{K}=10$, $\varepsilon = 0.10$). (A) Survival time and $\hat{\bm H}$. (B) ground truth $\bm W$, note that $\bm{W}_{[1,2]}$ associated with better prognosis (longer survival time), $\bm{W}_{[K-1,K]}$ associated with worse prognosis. (C) $\Tilde{\bm W}$ and hierarchical agglomerative clustering results (highlighted by most distinct colors, but did not relate with colors in (B) and (D)) with $\hat{K}$ number of clusters. Columns of $\Tilde{\bm W}$ and rows of $\hat{\bm H}$ were sorted in ascending order of $\hat{\beta}$. Cluster with highest mean value on the smallest $\hat{\beta}_{1}$ and cluster with highest mean value on the largest $\hat{\beta}_{K}$ were highlighted with black rectangle and \textcolor{magenta}{magenta} rectangle. (D) Ground truth labels in plot B with row permutation according to the hierarchical clustering result. In this figure, C-Index $=0.9996$ with accuracy $=0.8765$, F-1 score $=0.7050$, precision $=0.7748$, recall $=0.6721$, relative error $=10.9669\%$, and running time $=1.3064$ seconds.}
  \label{fig:simulation_3}
\end{center}
\end{figure*}

From Table \ref{table:main_univariate_X_noise=0} we observed that the proposed CoxNMF achieved highest C-Index (which was used for model selection) consistently among all $K$ in univariate underlying features simulation with $\varepsilon=0$. We also reported all univariate/multivariate simulation results with $\varepsilon\in\{0,0.05,0.10\}$ in Table \ref{table:main_univariate_X_noise=0.05}, \ref{table:main_univariate_X_noise=0.10}, \ref{table:main_multivariate_X_noise=0}, \ref{table:main_multivariate_X_noise=0.05}, and \ref{table:main_multivariate_X_noise=0.10}. The corresponding accuracy and relative error are also leads other models among all experiments: The relationships between C-Index and accuracy was reported in Figure \ref{fig:simulation_CI_vs_acc}, and the relationships between C-Index and relative error was reported in Figure \ref{fig:simulation_CI_vs_rerror}.
Two simulation results where $K=\hat{K}=10, \varepsilon=0.10$ was presented for univariate experiment (Figure \ref{fig:simulation_2}, C-Index $=0.9996$, accuracy $=0.9520$) and multivariate experiment (Figure \ref{fig:simulation_3}, C-Index $=0.9996$, accuracy $=0.8765$).
The complete univariate/multivariate simulation results including $\hat{K}$, relative errors, F-1 score, precision, recall, running time in seconds, etc. were further presented in Table \ref{table:supplement_univariate} and \ref{table:supplement_multivariate}.

In simulation results, CoxNMF demonstrated its superiority by reaching nearly 100\% C-Index, and achieved higher accuracy, F-1 score, etc. among all $K$ and $\varepsilon$, while the relative error is also competitive to others. We also found the proposed CoxNMF was especially robust to the noise $\bm E$ (when $\varepsilon$ is higher). The CoxNMF algorithm is also efficient with regards to the running time (Table \ref{table:supplement_univariate} and \ref{table:supplement_multivariate}). The quantitative results suggested that CoxNMF can accurately identify survival-related gene clusters.

\begin{figure}[t]
\begin{center}
  \includegraphics[width=\linewidth]{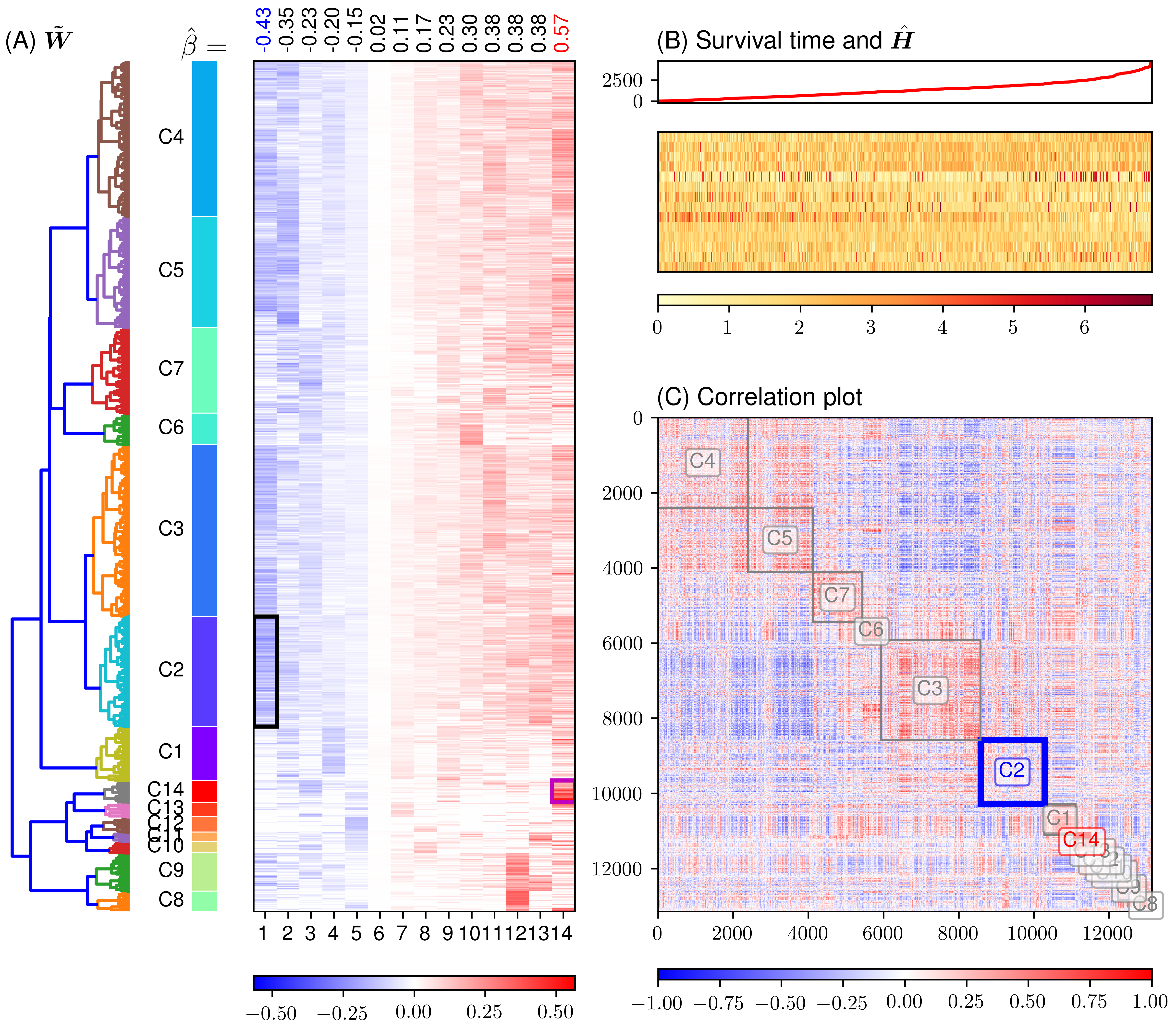}
  \caption{%
    Experimental results on Kidney Renal Clear Cell Carcinoma (KIRC). (A) hierarchical agglomerative clustering with $\hat{K}=14$ labels and the derived $\Tilde{\bm W}$ sorted by $\hat{\beta}$ in columns. (B) Survival time, and the corresponding $\hat{\bm H}$ sorted by survival time in columns, and sorted by $\hat{\beta}$ in rows. (C) Spearman's rank correlation plot of $\bm X$. Cluster labels are highlighted and resides at the block diagonal. Rectangles in \textcolor{blue}{blue} and \textcolor{red}{red} colors indicate the true location on $\bm X$ with respect to the clusters which are positively/negatively associated with survival, respectively.
    }
  \label{fig:KIRC}
\end{center}
\end{figure}

\subsection*{Human Cancer Gene Expressions}

To improve precision health and cancer treatments, we are particularly interested in discovering novel gene clusters behind the gene expression matrix and the corresponding survival information. The goal of discovering latent cancer gene interaction groups can help biologist reveal gene functions, setup biological experiments, or help develop drugs based on targeted genes. In this paper, two popular cancer types were used as demonstration to perform the CoxNMF, which are Kidney Renal Clear Cell Carcinoma (KIRC), and Lung Adenocarcinoma (LUAD).

\begin{figure*}[t]
\begin{center}
  \includegraphics[width=\linewidth]{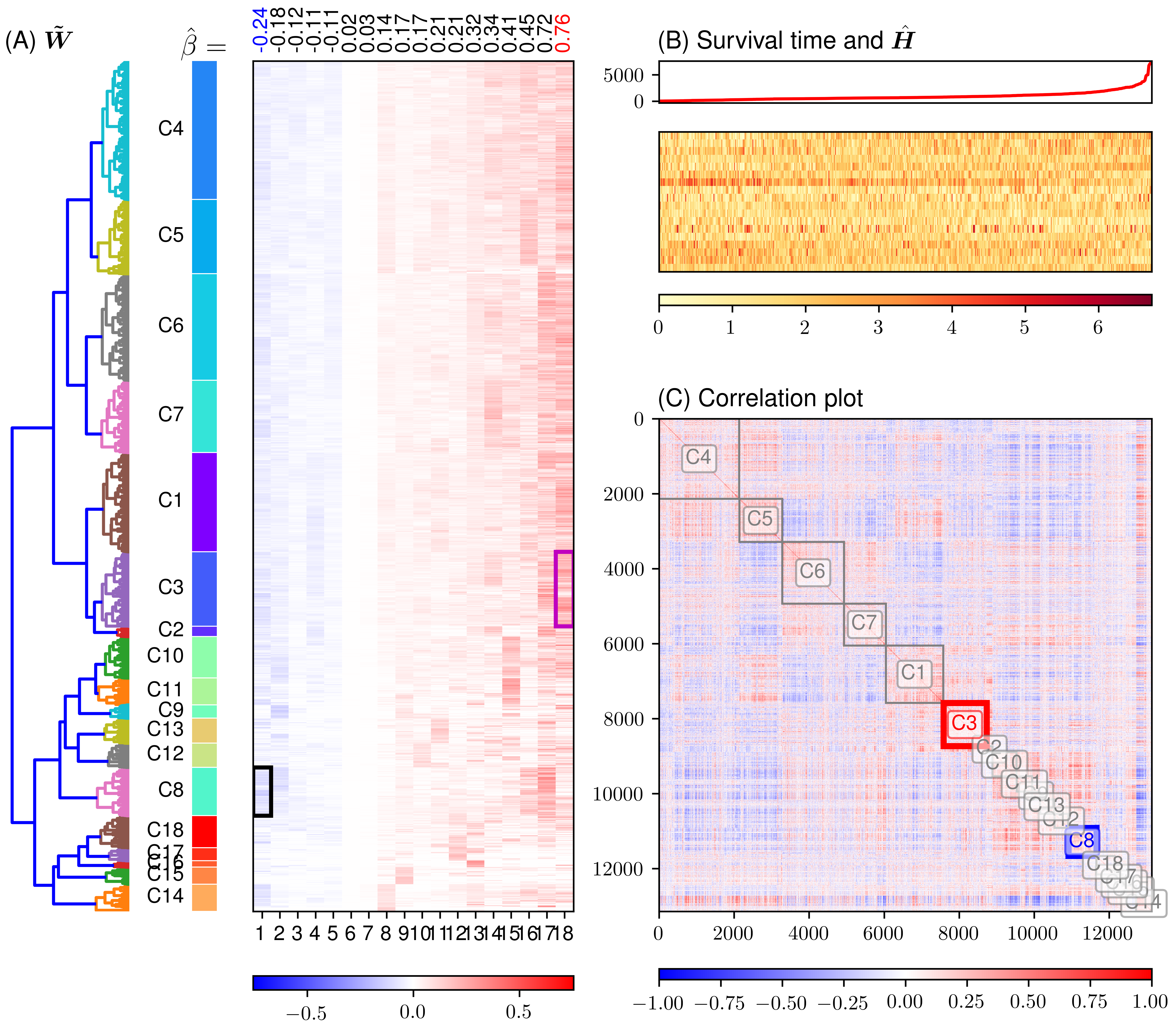}
  \caption{%
    Experimental results on Lung Adenocarcinoma (LUAD). (A) hierarchical agglomerative clustering with $\hat{K}=18$ labels and the derived $\Tilde{\bm W}$ sorted by $\hat{\beta}$ in columns. (B) Survival time, and the corresponding $\hat{\bm H}$ sorted by survival time in columns, and sorted by $\hat{\beta}$ in rows. (C) Spearman's rank correlation plot of $\bm X$. Cluster labels are highlighted and resides at the block diagonal. Rectangles in \textcolor{blue}{blue} and \textcolor{red}{red} colors indicate the true location on $\bm X$ with respect to the clusters which are positively/negatively associated with survival, respectively.
    }
  \label{fig:LUAD}
\end{center}
\end{figure*}

\begin{figure*}[h]
\begin{center}
  \includegraphics[width=0.4\linewidth]{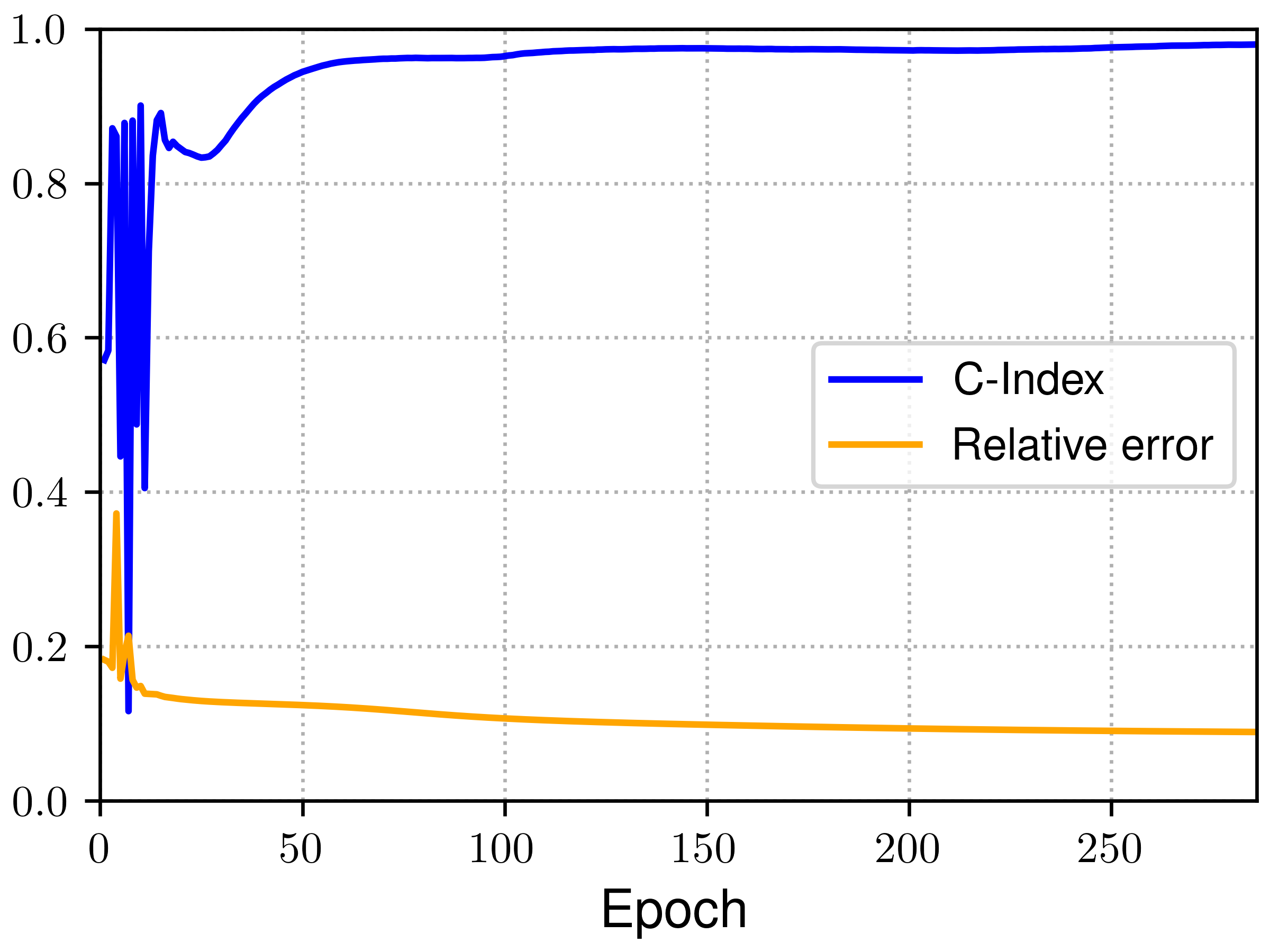}
  \caption{%
    Convergence of CoxNMF in Kidney Renal Clear Cell Carcinoma (KIRC) with $\hat{K}=14$. X-axes were truncated at the point where CoxNMF converged. Convergence may be achieved before the maximum iteration $M=500$. \textcolor{blue}{Blue line: C-Index}, \textcolor{orange}{Orange line: relative error}.
    }
  \label{fig:convergence_KIRC}
\end{center}
\end{figure*}

\begin{figure*}[h]
\begin{center}
  \includegraphics[width=0.4\linewidth]{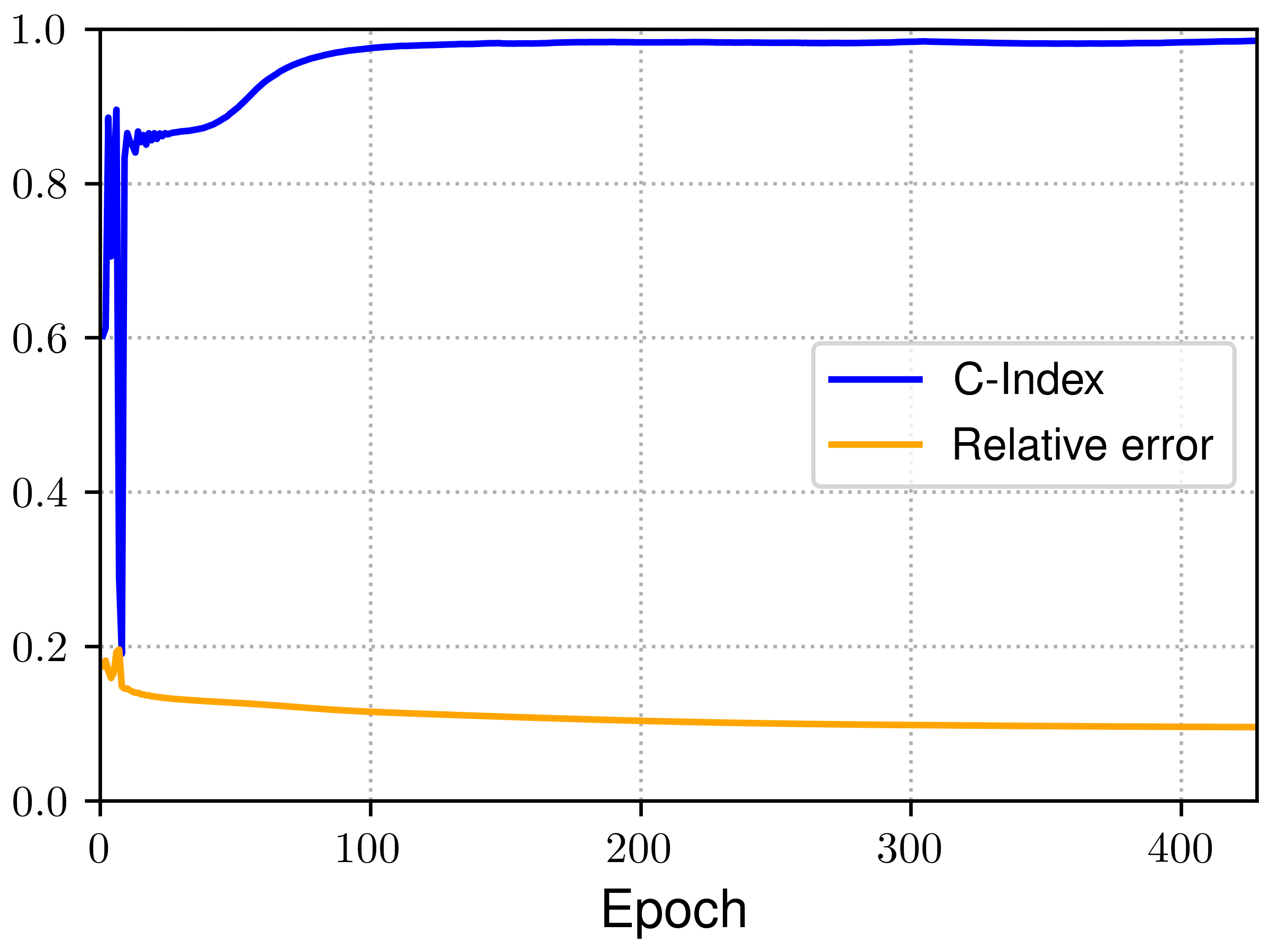}
  \caption{%
    Convergence of CoxNMF in Lung Adenocarcinoma (LUAD) with $\hat{K}=18$. X-axes were truncated at the point where CoxNMF converged. Convergence may be achieved before the maximum iteration $M=500$. \textcolor{blue}{Blue line: C-Index}, \textcolor{orange}{Orange line: relative error}.
    }
  \label{fig:convergence_LUAD}
\end{center}
\end{figure*}

Gene expression data (mRNA-seq) was downloaded from Broad GDAC Firehose (\url{https://gdac.broadinstitute.org/}). Since gene expressions remained a considerable amount of noises, 20\% of genes with lowest expression mean and 20\% of genes with lowest expression variance were excluded \cite{huang2019salmon}. All expressions were normalized in $\log_2$ scale: $\bm X = \log_2(\bm X+1)$ \cite{huang2019salmon}. We ended up with $P=$ 13,140 genes. In the experiments, $N=533$ patients for KIRC, and $N=507$ patients for LUAD. We searched $\hat{K}$ from 5 to 25 with step size $= 1$, and determined the hyper-parameters using the same criteria in the simulation study. The optimal $\hat{K}$ was determined where the relative error $<10\%$ meanwhile silhouette score is highest. We found $\hat{K}=14$ for KIRC, and $\hat{K}=18$ for LUAD. The detailed optimization results $\Tilde{\bm W}$ and $\hat{\bm H}$ for KIRC and LUAD were reported in Figure \ref{fig:KIRC} and \ref{fig:LUAD}. Besides, Spearman's rank correlation coefficient matrices were reported in Figure \ref{fig:KIRC}C and \ref{fig:LUAD}C, organized by the associated clusters concord/discord with the survival, suggested the effectiveness of low-rank reorganization in helping find survival associated clusters which is hard to exploit from the original data $\bm X$. Convergence plots for KIRC and LUAD were reported in Figure \ref{fig:convergence_KIRC} and \ref{fig:convergence_LUAD}, respectively.

By performing gene ontology (GO) analysis with ToppGene analysis suite \cite{chen2009toppgene}, certain GO biological process terms were exploited. The measurement of the enrichment results was based on $P$-value using the hypergeometric distribution \cite{boyle2004go}. A smaller $P$-value indicates a more significant association of gene cluster to a particular GO term. Top 3 GO terms with smallest $P$-value were reported. Among real human cancer data experiments, we investigated the gene clusters which have the strongest signal to the survival (assumed $\tau_1=\tau_2=1$).

For KIRC, identified gene cluster $C2$ was associated with better prognosis, while $C14$ behaved vice versa (Table \ref{table:KIRC}, Figure \ref{fig:KIRC}). From the results, we verified that important biological processes such as mRNA metabolic process ($P$-value $=1.244\times10^{-90}$), establishment of protein localization to endoplasmic reticulum (ER) ($P$-value $=1.982\times10^{-68}$), and protein targeting to ER ($P$-value $=1.353\times10^{-64}$) are associated with longer survival time, which verified the importance of ER in kidney cancer survival \cite{qiu2015hif2alpha}. Whereas for gene cluster $C14$, we found a strong association of worse survival prognosis with activation of GTPase activity ($P$-value $=1.698\times10^{-25}$), regulation of cilium assembly ($P$-value $=3.77\times10^{-23}$), and positive regulation of nucleic acid-templated transcription ($P$-value $=8.18\times10^{-17}$), echos several existed research findings \cite{wang2018regulation, liu2019association}.
For LUAD, identified gene cluster $C8$ was associated with better prognosis and is highly enriched with genes for activation of GTPase activity ($P$-value$=1.25\times 10^{-17}$) and sialylation ($P$-value$=1.85\times 10^{-16}$), while $C3$ behaved oppositely and is highly enriched with genes for organophosphate metabolic process ($P$-value$=6.10\times 10^{-71}$) as reported in Table \ref{table:LUAD} and Figure \ref{fig:LUAD}. Detailed enrichment analysis results for KIRC and LUAD were further elaborated in \textbf{Appendix}.
These findings demonstrated that CoxNMF can unravel survival associated gene clusters precisely, which can greatly help researchers identify cancer-specific survival-related gene modules as well as critical gene signatures.

\section{Conclusion}

In this paper, a novel algorithm CoxNMF was proposed by simultaneously learning the non-negative matrix factorization and the Cox proportional hazards regression. We designed the novel objective function and update rules for CoxNMF. To the best of our knowledge, this is the first work that performed non-negative matrix factorization and clustering driven by survival regression, which is accomplished by jointly optimization of the Frobenius norm and partial log likelihood. The proposed algorithm successfully demonstrated its superiority of identifying survival-associated gene clusters than other algorithms across thirty-six different synthetic data. The experiments conducted on human cancer datasets helped unravel latent gene clusters which reflect rich biological interpretations, achieved the goal of understanding and interpretation of high-dimensional biological data in precision health.

\section{Ethics Statement}
The proposed algorithm can help researchers discover critical genes associated with survival. When the important gene clusters are identified with the association to the survival, it is recommended to conduct further wet-lab experiments such as cancer cell line in order to validate the discovered biomarkers/genes.

\clearpage
\nolinenumbers
\section{Appendix}
\beginsupplement
\subsection*{Definition of Updating $\bm H$ for CoxNMF Algorithm}

\begin{definition}
\label{def:1}
We define the Equation \ref{eqn:CoxNMF_multiplicative_update_rule_H} to update $\bm H$ during CoxNMF updating
\begin{equation}
    \small
    \begin{aligned}
    \bm{H}^{(iter+1)}_{i,j} &\leftarrow \left(\bm{H}^{(iter)}_{i,j} + \frac{\alpha}{2} \text{max}\left\{ \bm{0},  \frac{\partial\ell_{\bm{H}^{(iter)},\beta}(C,Y)}{\partial\bm{H}^{(iter)}}\right\} \right) \\
    &\odot
\frac{{\bm{W}^{(iter+1)}_{i,j}}^T\bm{X}}{{\bm{W}^{(iter+1)}_{i,j}}^T\bm{W}^{(iter+1)}_{i,j}\bm{H}^{(iter)}_{i,j}},
    \end{aligned}
\end{equation}
where
\begin{equation}
    \small
    \begin{aligned}
        \frac{\partial\ell_{\bm{H},\beta}(C,Y)}{\partial\bm{H}} &= \begin{bmatrix}
        \frac{\partial\ell_{\bm{H},\beta}(C,Y)}{\partial\bm{H}_{1,1}} & \frac{\partial\ell_{\bm{H},\beta}(C,Y)}{\partial\bm{H}_{1,2}} & \cdots & \frac{\partial\ell_{\bm{H},\beta}(C,Y)}{\partial\bm{H}_{1,N}}\\
        \frac{\partial\ell_{\bm{H},\beta}(C,Y)}{\partial\bm{H}_{2,1}} & \frac{\partial\ell_{\bm{H},\beta}(C,Y)}{\partial\bm{H}_{2,2}} & \cdots & \frac{\partial\ell_{\bm{H},\beta}(C,Y)}{\partial\bm{H}_{2,N}}\\
        \vdots & \vdots & \ddots & \vdots\\
        \frac{\partial\ell_{\bm{H},\beta}(C,Y)}{\partial\bm{H}_{K,1}} & \frac{\partial\ell_{\bm{H},\beta}(C,Y)}{\partial\bm{H}_{K,2}} & \cdots & \frac{\partial\ell_{\bm{H},\beta}(C,Y)}{\partial\bm{H}_{K,N}}\\
        \end{bmatrix}\\
        &= \begin{bmatrix}
        \underbrace{\left(C_r\beta - \sum\limits^{N}_{s=r}C_s\frac{\mathds{1}_{(Y_r\geq Y_s)}\beta\text{exp}(\beta^T\bm{H}_r)}{\sum_{j:Y_j\geq Y_s}\text{exp}(\beta^T\bm{H}_j)}\right)}_\text{ $K\times 1$ vector which repeats $N$ times for $r=1,2,\cdots,N$.}
        \end{bmatrix}
    \end{aligned}
\end{equation}
and $\mathds{1}_{(Y_i\geq Y_s)} = 
\begin{cases}
  1 & \text{if } Y_i\geq Y_s \\
  0 & \text{otherwise}
\end{cases}$ is the indicator function.
\end{definition}

\begin{proof}
since the partial derivative $\frac{\partial\left(\norm{\bm{X}-\bm W \bm H}^2_F-\alpha\ell_{\bm{H},\beta}(C,Y)\right)}{\partial\bm{H}}$ is
\begin{equation}
    \small
    \begin{aligned}
        \frac{\partial\left(\norm{\bm{X}-\bm W \bm H}^2_F-\alpha\ell_{\bm{H},\beta}(C,Y)\right)}{\partial\bm{H}} = &-2\bm{W}^T\bm{X} + 2\bm{W}^T\bm{W}\bm{H}\\
        &- \alpha\frac{\partial\ell_{\bm{H},\beta}(C,Y)}{\partial\bm{H}}.
    \end{aligned}
\end{equation}
Thus we have the update rule for $\bm{H}$:
\begin{equation}
\small
\label{eqn:Cox_NMF_update_rule_H_detailed_derivation}
\begin{aligned}
    \bm{H} &\leftarrow \bm{H} - \eta_{\bm{H}} \odot \frac{\partial\left(\norm{\bm{X}-\bm W \bm H}^2_F-\alpha\ell_{\bm{H},\beta}(C,Y)\right)}{\partial\bm{H}}\\
    &\leftarrow \bm{H} - \eta_{\bm{H}} \odot \left[-2\bm{W}^T\bm{X} + 2\bm{W}^T\bm{W}\bm{H} - \alpha\frac{\partial\ell_{\bm{H},\beta}(C,Y)}{\partial\bm{H}}\right]\\
    &\leftarrow \bm{H} - \eta_{\bm{H}} \odot \left[-2\bm{W}^T\bm{X} + 2\bm{W}^T\bm{W}\bm{H}\right] + \alpha\eta_{\bm{H}} \odot \frac{\partial\ell_{\bm{H},\beta}(C,Y)}{\partial\bm{H}}\\
    &\leftarrow \bm{H} + \eta_{\bm{H}} \odot \left[2\bm{W}^T\bm{X} - 2\bm{W}^T\bm{W}\bm{H}\right] \\ &\;\;\;\;\;\;\;\;\;\;  + \alpha\eta_{\bm{H}} \odot
    \Bigg[
    \left(C_1\beta - \sum\limits^{N}_{s=1}C_s\frac{\mathds{1}_{(Y_1\geq Y_s)}\beta\text{exp}(\beta^T\bm{H}_1)}{\sum_{j:Y_j\geq Y_s}\text{exp}(\beta^T\bm{H}_j)}\right) \cdots \\
    & \;\;\;\;\;\;\;\;\;\;\;\;\;\;\;\;\;\;\;\; \left(C_N\beta - \sum\limits^{N}_{s=1}C_s\frac{\mathds{1}_{(Y_n\geq Y_s)}\beta\text{exp}(\beta^T\bm{H}_N)}{\sum_{j:Y_j\geq Y_s}\text{exp}(\beta^T\bm{H}_j)}\right)\Bigg].
\end{aligned}
\end{equation}

Given the same $\eta_{\bm{H}}=\frac{\bm{H}}{2\bm{W}^T\bm{W}\bm{H}}$, as long as we project $C_r\beta - \sum\limits^{N}_{s=r}C_s\frac{\mathds{1}_{(Y_r\geq Y_s)}\beta\text{exp}(\beta^T\bm{H}_r)}{\sum_{j:Y_j\geq Y_s}\text{exp}(\beta^T\bm{H}_j)}$ into the first orthant of $K$-dimensional space, make sure all the elements in $C_r\beta - \sum\limits^{N}_{s=r}C_s\frac{\mathds{1}_{(Y_r\geq Y_s)}\beta\text{exp}(\beta^T\bm{H}_r)}{\sum_{j:Y_j\geq Y_s}\text{exp}(\beta^T\bm{H}_j)} \geq 0$, then we can guarantee the updating is non-negative with respect to $\bm{H}$.

According this projection rule, we get
\begin{equation}
    \small
    \bm{H}_{i,j} \leftarrow \left(\bm{H}_{i,j} + \frac{\alpha}{2} \text{max}\left\{ \bm{0},  \frac{\partial\ell_{\bm{H},\beta}(C,Y)}{\partial\bm{H}}\right\} \right) \odot
\frac{{\bm{W}_{i,j}}^T\bm{X}}{{\bm{W}_{i,j}}^T\bm{W}_{i,j}\bm{H}_{i,j}},
\end{equation}
where
\begin{equation}
    \small
    \begin{aligned}
    \frac{\partial\ell_{\bm{H},\beta}(C,Y)}{\partial\bm{H}} = \Bigg[ &\left(C_1\beta - \sum\limits^{N}_{s=1}C_s\frac{\mathds{1}_{(Y_1\geq Y_s)}\beta\text{exp}(\beta^T\bm{H}_1)}{\sum_{j:Y_j\geq Y_s}\text{exp}(\beta^T\bm{H}_j)}\right) \cdots \\
    &  \left(C_N\beta - \sum\limits^{N}_{s=1}C_s\frac{\mathds{1}_{(Y_N\geq Y_s)}\beta\text{exp}(\beta^T\bm{H}_N)}{\sum_{j:Y_j\geq Y_s}\text{exp}(\beta^T\bm{H}_j)}\right)\Bigg].
    \end{aligned}
\end{equation}
\end{proof}

\definecolor{bgcolor}{RGB}{255,246,230}
\begin{mdframed}[backgroundcolor=bgcolor,rightline=false,leftline=false]
    \textbf{Example}\\
    \small{
An example data is provided to better understand the Equation \ref{eqn:Cox_NMF_update_rule_H_detailed_derivation}. Suppose our $\bm{H}$ is with dimension $K$ by 4 (4 patients), we have survival time $Y = [1, 3, 2, 4]$ and survival event $C = [1, 0, 1, 1]$. Then the partial derivative $\frac{\partial\ell_{\bm{H},\beta}(C,Y)}{\partial\bm{H}}$ becomes
\begin{equation}
\tiny
\begin{aligned}
    &\frac{\partial\ell_{\bm{H},\beta}(C,Y)}{\partial\bm{H}} =\frac{\partial}{\partial\bm{H}}\sum_{i:C_i = 1}\big(\beta^T\bm{H}_i - \text{log}\big( \sum_{j:Y_j\geq Y_i} \text{exp}(\beta^T\bm{H}_j) \big)\big)\\
    &=\frac{\partial}{\partial\bm{H}}\bigg(\beta^T\bm{H}_1 - \text{log}\big[\text{exp}(\beta^T\bm{H}_1 ) + \text{exp}(\beta^T\bm{H}_2 ) \\
    &\;\;\;\;\;\;\;\;\;\; + \text{exp}(\beta^T\bm{H}_3 ) + \text{exp}(\beta^T\bm{H}_4 )\big] \\
    &\;\;\;\;\;\;\;\;\;\;+ 0\cdot\big[\beta^T\bm{H}_2 - \text{log}\big[\text{exp}(\beta^T\bm{H}_2) + \text{exp}(\beta^T\bm{H}_4)\big]\big] \\
    &\;\;\;\;\;\;\;\;\;\;+ \beta^T\bm{H}_3 - \text{log}\big[\text{exp}(\beta^T\bm{H}_2) + \text{exp}(\beta^T\bm{H}_3) + \text{exp}(\beta^T\bm{H}_4)\big]\\
    &\;\;\;\;\;\;\;\;\;\;+ \beta^T\bm{H}_4 - \text{log}\big[\text{exp}(\beta^T\bm{H}_4 )\big]\bigg)\\
    &= \frac{\partial}{\partial\bm{H}}\bigg(\beta^T\bm{H}_1 - \text{log}\big[\text{exp}(\beta^T\bm{H}_1 ) + \text{exp}(\beta^T\bm{H}_2 ) \\
    &\;\;\;\;\;\;\;\;\;\;+ \text{exp}(\beta^T\bm{H}_3 ) + \text{exp}(\beta^T\bm{H}_4 )\big] \\
    &\;\;\;\;\;\;\;\;\;\;+ \beta^T\bm{H}_3 - \text{log}\big[\text{exp}(\beta^T\bm{H}_2) + \text{exp}(\beta^T\bm{H}_3) + \text{exp}(\beta^T\bm{H}_4)\big]\bigg).
\end{aligned}
\end{equation}
That is,
\begin{equation}
\tiny
\begin{aligned}
    \frac{\partial\ell_{\bm{H},\beta}(C,Y)}{\partial\bm{H}_1} &= \beta - \\
    & \frac{\beta\text{exp}(\beta^T\bm{H}_1)}{\text{exp}(\beta^T\bm{H}_1 ) + \text{exp}(\beta^T\bm{H}_2 ) + \text{exp}(\beta^T\bm{H}_3 ) + \text{exp}(\beta^T\bm{H}_4 )},
\end{aligned}
\end{equation}
\begin{equation}
\tiny
\begin{aligned}
    \frac{\partial\ell_{\bm{H},\beta}(C,Y)}{\partial\bm{H}_2} &= - \frac{\beta\text{exp}(\beta^T\bm{H}_2)}{\text{exp}(\beta^T\bm{H}_1 ) + \text{exp}(\beta^T\bm{H}_2 ) + \text{exp}(\beta^T\bm{H}_3 ) + \text{exp}(\beta^T\bm{H}_4 )}\\
    &\;\;\;\; - \frac{\beta\text{exp}(\beta^T\bm{H}_2)}{\text{exp}(\beta^T\bm{H}_2 ) + \text{exp}(\beta^T\bm{H}_3 ) + \text{exp}(\beta^T\bm{H}_4 )},
\end{aligned}
\end{equation}
\begin{equation}
\tiny
\begin{aligned}
    \frac{\partial\ell_{\bm{H},\beta}(C,Y)}{\partial\bm{H}_3} &= \beta + \frac{\beta\text{exp}(\beta^T\bm{H}_3)}{\text{exp}(\beta^T\bm{H}_2 ) + \text{exp}(\beta^T\bm{H}_3 ) + \text{exp}(\beta^T\bm{H}_4 )} \\
    &- \frac{\beta\text{exp}(\beta^T\bm{H}_3)}{\text{exp}(\beta^T\bm{H}_1 ) + \text{exp}(\beta^T\bm{H}_2 ) + \text{exp}(\beta^T\bm{H}_3 ) + \text{exp}(\beta^T\bm{H}_4 )},
\end{aligned}
\end{equation}
\begin{equation}
\tiny
\begin{aligned}
    \frac{\partial\ell_{\bm{H},\beta}(C,Y)}{\partial\bm{H}_4} &= - \frac{\beta\text{exp}(\beta^T\bm{H}_4)}{\text{exp}(\beta^T\bm{H}_1 ) + \text{exp}(\beta^T\bm{H}_2 ) + \text{exp}(\beta^T\bm{H}_3 ) + \text{exp}(\beta^T\bm{H}_4 )}\\
    &\;\;\;\; - \frac{\beta\text{exp}(\beta^T\bm{H}_4)}{\text{exp}(\beta^T\bm{H}_2 ) + \text{exp}(\beta^T\bm{H}_3 ) + \text{exp}(\beta^T\bm{H}_4 )}.
\end{aligned}
\end{equation}
And for each patient $r$, the partial derivative $\frac{\partial\ell_{\bm{H},\beta}(C,Y)}{\partial\bm{H}_r}$ is
\begin{equation}
    \tiny
    C_r\beta - \sum\limits^{N}_{s=1}C_s\frac{\mathds{1}_{(Y_r\geq Y_s)}\beta\text{exp}(\beta^T\bm{H}_r)}{\sum_{j:Y_j\geq Y_s}\text{exp}(\beta^T\bm{H}_j)}.
\end{equation}
}

\end{mdframed}

\subsection*{Proof of Time Complexity}

\begin{theorem}
\label{thm:time_complexity}
The time complexity of CoxNMF update for one iteration consists of Equation \ref{eqn:CoxNMF_multiplicative_update_rule_W}, \ref{eqn:CoxNMF_multiplicative_update_rule_beta}, and \ref{eqn:CoxNMF_multiplicative_update_rule_H} is $O(PNK+N^2K+K^2\max(P,N))$.
\end{theorem}

\begin{proof}
For Equation \ref{eqn:CoxNMF_multiplicative_update_rule_W}, since in general $K < \min(P,N)$, calculating the denominator in the form of $(\bm W \bm H)\bm H^T$ would cost $O(PNK)$, calculating $\bm W(\bm H\bm H^T)$ costs $O(\max(P,N)K^2)$, so the latter method is better \cite{lin2007projected}. For the numerator, $\bm X \bm H^T$ costs $O(PNK)$. Thus, Equation \ref{eqn:CoxNMF_multiplicative_update_rule_W} takes $O(PNK + \max(P,N)K^2 + PK)$ operations.

For Equation \ref{eqn:CoxNMF_multiplicative_update_rule_beta}, since multivariate Newton-Raphson method (number of independent variables = $K$) was used to solve MPLE problem \cite{gonzalez2008estimation}, and notice that we can pre-calculate some common structures. For example, the indication matrix $\sum_i{\sum_{j: Y_j \geq Y_i}}$ will cost $O(N^2)$, the $\exp(\beta^T\bm{H})$ will cost $O(NK^2)$. 
Thus, we can take $O(NK^2+N^2+N+K^2N+NK+K^2)=O(NK^2+N^2)$ operations to get $\mathcal{H}$. Similarly, we can get $g$ with $O(NK^2+N^2+NK)$ operations. Since the inverse of Hessian matrix $\mathcal{H}^{-1}$ can be $O(K^2\log(K))$ theoretically, and $\mathcal{H}^{-1}g$ will cost $O(K^2)$, thus the time complexity for Equation \ref{eqn:CoxNMF_multiplicative_update_rule_beta} is $O(N^2+K^2(N+\log(K)))$.

For Equation \ref{eqn:CoxNMF_multiplicative_update_rule_H}, since the partial derivative of the partial log likelihood $\ell_{\bm{H},\beta}(C,Y)$ with respect to $\bm H$ in Equation \ref{eqn:CoxNMF_multiplicative_update_rule_H_detail} costs $O(NK^2+N^2+N^2K+N^2) = O(NK^2+N^2K)$ operations, thus the Equation \ref{eqn:CoxNMF_multiplicative_update_rule_H} takes $O(PNK + \max(P,N)K^2 + NK^2+N^2K + NK)$ operations.

Thus, the time complexity of CoxNMF update for one iteration consists of Equation \ref{eqn:CoxNMF_multiplicative_update_rule_W}, \ref{eqn:CoxNMF_multiplicative_update_rule_beta}, and \ref{eqn:CoxNMF_multiplicative_update_rule_H} is $O(PNK + \max(P,N)K^2 + PK + N^2+K^2(N+\log(K)) + PNK + \max(P,N)K^2 + NK^2+N^2K + NK) = O(PNK+N^2K+K^2\max(P,N))$.

\end{proof}

\subsection*{Proof of Space Complexity}

\begin{theorem}
\label{thm:space_complexity}
The space complexity of CoxNMF update for one iteration consists of Equation \ref{eqn:CoxNMF_multiplicative_update_rule_W}, \ref{eqn:CoxNMF_multiplicative_update_rule_beta}, and \ref{eqn:CoxNMF_multiplicative_update_rule_H} is $O(PN+N^2)$.
\end{theorem}

\begin{proof}

For Equation \ref{eqn:CoxNMF_multiplicative_update_rule_W}, since in general $K < \min(P,N)$, the denominator in the form of $(\bm W \bm H)\bm H^T$ would consume $O(PK+NK+K^2)=O(PK+NK)$ spaces, the numerator $\bm X \bm H^T$ consume $O(PN+NK)$ spaces. Thus, Equation \ref{eqn:CoxNMF_multiplicative_update_rule_W} consume $O(PN+PK+NK) = O(PN)$ spaces.

For Equation \ref{eqn:CoxNMF_multiplicative_update_rule_beta}, notice that the common structure $\exp(\beta^T\bm{H})$ consume $O(KN)$ spaces, and the indication matrix $\sum_i{\sum_{j: Y_j \geq Y_i}}$ will consume $O(N^2)$ spaces, thus it consumes $O(N^2+NK+K^2) = O(N^2)$ spaces to get $\mathcal{H}$. Similarly, we can get $g$ with $O(K+KN)$ spaces consumed. Thus, the space complexity for Equation \ref{eqn:CoxNMF_multiplicative_update_rule_beta} is $O(KN+N^2+K+KN) = O(N^2)$.

For Equation \ref{eqn:CoxNMF_multiplicative_update_rule_H}, since the partial derivative of the partial log likelihood $\ell_{\bm{H},\beta}(C,Y)$ with respect to $\bm H$ in Equation \ref{eqn:CoxNMF_multiplicative_update_rule_H_detail} consumes $O(N^2+KN)$ spaces, thus the Equation \ref{eqn:CoxNMF_multiplicative_update_rule_H} consumes $O(PN+N^2)$ operations.

Thus, the space complexity of CoxNMF update for one iteration consists of Equation \ref{eqn:CoxNMF_multiplicative_update_rule_W}, \ref{eqn:CoxNMF_multiplicative_update_rule_beta}, and \ref{eqn:CoxNMF_multiplicative_update_rule_H} is $O(PN+N^2)$.
\end{proof}

\subsection*{Human Cancer Gene Expressions (cont.)}
For Lung Adenocarcinoma (LUAD), identified gene cluster $C8$ was associated with better prognosis, while $C3$ behaved in contrast (Table \ref{table:LUAD}). From the results, we found intracellular protein transport ($P$-value $=1.254\times10^{-22}$), which could be a mediator for tumor suppressor activity \cite{sieger2004tumor}, was the most significant GO terms concord with the survival. Interestingly, the activation of GTPase acitivty ($P$-value $=1.489\times10^{-17}$), which was ranked top for discording KIRC survival, is now ranked top for concording LUAD survival. For the worse LUAD survival, top ranked GO terms are all metabolic processes. For example, strong evidences were found from previous literature which clearly stated that exposed to organophosphate ($P$-value $=6.104\times10^{-71}$) can lead to lung cancer risk \cite{jones2015incidence}.

\subsection{Supplementary Tables and Figures}

\begin{table}[h]
    \small
    \centering
    \begin{tabular}{c l}
    \hline
    Symbols & Descriptions and data types\\
    \hline
    $P$ & Number of features/covariates. Scalar value.\\
    $N$ & Number of patients/samples. Scalar value.\\
    $K$ & Low-rank dimensions. Scalar value.\\
    $\bm X$ & Input matrix. $P\times N$ non-negative matrix.\\
    $\bm W$ & Basis matrix. $P\times K$ non-negative matrix.\\
    $\bm H$ & Coefficient matrix. $K\times N$ non-negative matrix.\\
    $\beta$ & Cox model parameters. $K\times 1$ vector.\\
    $\Tilde{\bm W}$ & Weighted basis matrix. $P\times K$ matrix.\\
    $Y$ & Survival times. $N\times 1$ vector.\\
    $C$ & Survival events. $N\times 1$ vector.\\
    $r$ & Risk score for C-Index calculation. Scalar value.\\
    $M$ & Number of the iterations. Scalar value.\\
    $iter$ & Current iteration. Scalar value.\\
    $\alpha$ & Step size for $\bm H$ in CoxNMF. Scalar value.\\
    $\gamma$ & L1 penalty ratio. Scalar value $\in [0,1]$.\\
    $\eta$ & Parameter for smoothed L1 norm. Scalar value. \\
    $\xi$ & Regularization weight. Scalar value.\\
    $\Psi$ & Indices set for Efron's method. Set.\\
    $\mathcal{H}$ & Hessian matrix. $K\times K$ matrix.\\
    $g$ & Gradient of log partial likelihood. $K\times 1$ vector.\\
    $m_j$ & Number of the death count at $Y_j$. Scalar value.\\
    $\tau_1$ & Number of worse prognosis bases in simulation.\\
    $\tau_2$ & Number of better prognosis bases in simulation.\\
    $L_i$ & Gene cluster indicator. $i=1,\cdots,P$.\\
    $\varepsilon$ & Parameter in exponential distribution. Scalar value.\\
    \hline
    \end{tabular}
    \caption{Symbol description.}
    \label{table:symbol_description}
\end{table}

\begin{table*}[t]
    \centering
    \small
    \begin{tabular}{llllllll}
    \toprule
             & $K$  &            7 &            8 &            9 &           10 &           11 &           12 \\
    Metrics & Model &              &              &              &              &              &              \\
    \midrule
    C-Index & TruncatedSVD &  0.8586$\pm$0.07 &  0.8403$\pm$0.07 &  0.8248$\pm$0.15 &  0.8786$\pm$0.05 &  0.8499$\pm$0.15 &  0.8079$\pm$0.20 \\
             & PCA &  0.8506$\pm$0.08 &  0.5594$\pm$0.04 &  0.8329$\pm$0.15 &  0.9856$\pm$0.00 &  0.9884$\pm$0.00 &  0.7968$\pm$0.22 \\
             & SparsePCA &  0.9553$\pm$0.01 &  0.7628$\pm$0.09 &  0.9533$\pm$0.01 &  0.9524$\pm$0.01 &  0.7899$\pm$0.15 &  0.7594$\pm$0.18 \\
             & NNDSVD &  0.5646$\pm$0.03 &  0.5564$\pm$0.03 &  0.6129$\pm$0.05 &  0.5665$\pm$0.03 &  0.6010$\pm$0.05 &  0.6433$\pm$0.06 \\
             & FactorAnalysis &  0.9861$\pm$0.00 &  0.8296$\pm$0.13 &  0.9855$\pm$0.00 &  0.9855$\pm$0.00 &  0.9888$\pm$0.00 &  0.8092$\pm$0.20 \\
             & NMF (CD) &  0.8585$\pm$0.07 &  0.8398$\pm$0.07 &  0.8266$\pm$0.15 &  0.8780$\pm$0.05 &  0.8403$\pm$0.17 &  0.8084$\pm$0.20 \\
             & NMF (MU) &  0.7728$\pm$0.10 &  0.6923$\pm$0.12 &  0.6882$\pm$0.16 &  0.7044$\pm$0.07 &  0.6563$\pm$0.07 &  0.6498$\pm$0.12 \\
             & SNMF &  0.8575$\pm$0.07 &  0.8440$\pm$0.07 &  0.9854$\pm$0.00 &  0.8782$\pm$0.05 &  0.8505$\pm$0.15 &  0.8084$\pm$0.20 \\
             & CoxNMF &  \textbf{0.9997$\pm$0.00} &  \textbf{0.9998$\pm$0.00} &  \textbf{0.9998$\pm$0.00} &  \textbf{0.9996$\pm$0.00} &  \textbf{1.0000$\pm$0.00} &  \textbf{0.9993$\pm$0.00} \\
    \midrule
    Accuracy & TruncatedSVD &  0.7183$\pm$0.08 &  0.7012$\pm$0.06 &  0.7938$\pm$0.09 &  0.7888$\pm$0.05 &  0.8453$\pm$0.01 &  0.8497$\pm$0.05 \\
             & PCA &  0.7814$\pm$0.06 &  0.7853$\pm$0.05 &  0.8109$\pm$0.02 &  0.8932$\pm$0.03 &  0.9120$\pm$0.04 &  0.8905$\pm$0.02 \\
             & SparsePCA &  0.7111$\pm$0.04 &  0.7490$\pm$0.06 &  0.7927$\pm$0.06 &  0.8634$\pm$0.08 &  0.7756$\pm$0.12 &  0.8382$\pm$0.08 \\
             & NNDSVD &  0.7194$\pm$0.01 &  0.7932$\pm$0.06 &  0.7913$\pm$0.06 &  0.8300$\pm$0.05 &  0.8371$\pm$0.04 &  0.8750$\pm$0.04 \\
             & FactorAnalysis &  \textbf{0.9123$\pm$0.05} &  0.7550$\pm$0.01 &  0.9131$\pm$0.05 &  0.9284$\pm$0.04 &  \textbf{0.9353$\pm$0.05} &  0.8768$\pm$0.05 \\
             & NMF (CD) &  0.7197$\pm$0.07 &  0.7068$\pm$0.06 &  0.7764$\pm$0.06 &  0.7958$\pm$0.05 &  0.8396$\pm$0.01 &  0.8405$\pm$0.03 \\
             & NMF (MU) &  0.7640$\pm$0.10 &  0.7220$\pm$0.06 &  0.8211$\pm$0.07 &  0.8014$\pm$0.03 &  0.8291$\pm$0.01 &  0.8400$\pm$0.01 \\
             & SNMF &  0.7754$\pm$0.08 &  0.7927$\pm$0.08 &  \textbf{0.9140$\pm$0.06} &  0.8092$\pm$0.04 &  0.8409$\pm$0.02 &  0.8723$\pm$0.04 \\
             & CoxNMF &  0.8380$\pm$0.03 &  \textbf{0.8705$\pm$0.00} &  0.9056$\pm$0.04 &  \textbf{0.9294$\pm$0.05} &  0.9095$\pm$0.04 &  \textbf{0.9310$\pm$0.05} \\
    \bottomrule
    \end{tabular}
    \caption{Simulation results with univariate underlying features setup for $K=\{7,8,9,10,11,12\}$, $\varepsilon = 0.05$. $\hat{K}$ was searched around $K\pm\{0,1,2\}$ and was determined by highest silhouette score. Experiments repeat 5 times each with random seed $\in\{1,2,3,4,5\}$. Mean values $\pm$ standard deviations were reported, best performed mean values among models were highlighted in bold font.}
    \label{table:main_univariate_X_noise=0.05}
\end{table*}

\begin{table*}[t]
    \centering
    \small
    \begin{tabular}{llllllll}
    \toprule
             & $K$  &            7 &            8 &            9 &           10 &           11 &           12 \\
    Metrics & Model &              &              &              &              &              &              \\
    \midrule
    C-Index & TruncatedSVD &  0.7746$\pm$0.10 &  0.7477$\pm$0.07 &  0.7636$\pm$0.16 &  0.7799$\pm$0.08 &  0.7819$\pm$0.15 &  0.7651$\pm$0.19 \\
             & PCA &  0.5590$\pm$0.03 &  0.9651$\pm$0.01 &  0.5711$\pm$0.05 &  0.5711$\pm$0.03 &  0.7991$\pm$0.14 &  0.5758$\pm$0.03 \\
             & SparsePCA &  0.7317$\pm$0.12 &  0.7200$\pm$0.09 &  0.9430$\pm$0.01 &  0.9426$\pm$0.01 &  0.7580$\pm$0.14 &  0.7455$\pm$0.16 \\
             & NNDSVD &  0.5641$\pm$0.03 &  0.5568$\pm$0.03 &  0.5767$\pm$0.02 &  0.5676$\pm$0.04 &  0.5755$\pm$0.03 &  0.6383$\pm$0.05 \\
             & FactorAnalysis &  0.7686$\pm$0.11 &  0.7532$\pm$0.09 &  0.7655$\pm$0.15 &  0.8145$\pm$0.08 &  0.7828$\pm$0.16 &  0.7736$\pm$0.18 \\
             & NMF (CD) &  0.7735$\pm$0.10 &  0.7467$\pm$0.07 &  0.7626$\pm$0.16 &  0.7784$\pm$0.08 &  0.7812$\pm$0.15 &  0.7659$\pm$0.19 \\
             & NMF (MU) &  0.7328$\pm$0.09 &  0.6860$\pm$0.07 &  0.6604$\pm$0.15 &  0.6694$\pm$0.07 &  0.6631$\pm$0.07 &  0.6511$\pm$0.10 \\
             & SNMF &  0.7749$\pm$0.10 &  0.7554$\pm$0.08 &  0.7629$\pm$0.16 &  0.7798$\pm$0.08 &  0.7851$\pm$0.15 &  0.7740$\pm$0.17 \\
             & CoxNMF &  \textbf{0.9989$\pm$0.00} &  \textbf{0.9992$\pm$0.00} &  \textbf{0.9976$\pm$0.00} &  \textbf{0.9982$\pm$0.00} &  \textbf{0.9987$\pm$0.00} &  \textbf{0.9973$\pm$0.00} \\
    \midrule
    Accuracy & TruncatedSVD &  0.7140$\pm$0.07 &  0.6890$\pm$0.06 &  0.7844$\pm$0.08 &  0.7834$\pm$0.05 &  0.8173$\pm$0.01 &  0.8290$\pm$0.04 \\
             & PCA &  0.7514$\pm$0.04 &  0.8417$\pm$0.03 &  0.8169$\pm$0.02 &  0.8470$\pm$0.02 &  0.8735$\pm$0.02 &  0.8633$\pm$0.02 \\
             & SparsePCA &  0.7311$\pm$0.09 &  0.7358$\pm$0.07 &  0.8213$\pm$0.01 &  0.8280$\pm$0.08 &  0.7535$\pm$0.09 &  0.8028$\pm$0.08 \\
             & NNDSVD &  0.7194$\pm$0.02 &  0.7878$\pm$0.05 &  0.7871$\pm$0.01 &  0.8328$\pm$0.05 &  0.8229$\pm$0.02 &  0.8712$\pm$0.04 \\
             & FactorAnalysis &  0.7531$\pm$0.09 &  0.7540$\pm$0.05 &  0.8009$\pm$0.05 &  0.8226$\pm$0.05 &  0.8633$\pm$0.01 &  0.8722$\pm$0.05 \\
             & NMF (CD) &  0.7440$\pm$0.10 &  0.6847$\pm$0.06 &  0.7651$\pm$0.04 &  0.7854$\pm$0.05 &  0.8295$\pm$0.01 &  0.8233$\pm$0.03 \\
             & NMF (MU) &  0.7606$\pm$0.10 &  0.7000$\pm$0.05 &  0.7944$\pm$0.07 &  0.8022$\pm$0.03 &  0.8380$\pm$0.01 &  0.8443$\pm$0.00 \\
             & SNMF &  0.7760$\pm$0.10 &  0.7637$\pm$0.07 &  0.7991$\pm$0.07 &  0.8286$\pm$0.05 &  0.8382$\pm$0.01 &  0.8473$\pm$0.01 \\
             & CoxNMF &  \textbf{0.8557$\pm$0.05} &  \textbf{0.8850$\pm$0.04} &  \textbf{0.9058$\pm$0.04} &  \textbf{0.8888$\pm$0.04} &  \textbf{0.8922$\pm$0.01} &  \textbf{0.8997$\pm$0.02} \\
    \bottomrule
    \end{tabular}
    \caption{Simulation results with univariate underlying features setup for $K\in\{7,8,9,10,11,12\}$, $\varepsilon = 0.10$. $\hat{K}$ was searched around $K\pm\{0,1,2\}$ and was determined by highest silhouette score. Experiments repeat 5 times each with random seed $\in\{1,2,3,4,5\}$. Mean values $\pm$ standard deviations were reported, best performed mean values among models were highlighted in bold font.}
    \label{table:main_univariate_X_noise=0.10}
\end{table*}

\begin{table*}[t]
    \centering
    \small
    \begin{tabular}{llllllll}
    \toprule
             & $K$  &            7 &            8 &            9 &           10 &           11 &           12 \\
    Metrics & Model &              &              &              &              &              &              \\
    \midrule
    C-Index & TruncatedSVD &  \textbf{0.9997$\pm$0.00} &  \textbf{0.9993$\pm$0.00} &  0.9992$\pm$0.00 &  0.9989$\pm$0.00 &  0.9983$\pm$0.00 &  0.9979$\pm$0.00 \\
             & PCA &  \textbf{0.9997$\pm$0.00} &  \textbf{0.9993$\pm$0.00} &  0.9992$\pm$0.00 &  0.9988$\pm$0.00 &  0.9983$\pm$0.00 &  0.9891$\pm$0.01 \\
             & SparsePCA &  0.9631$\pm$0.00 &  0.9484$\pm$0.02 &  0.9567$\pm$0.01 &  0.9555$\pm$0.01 &  0.8141$\pm$0.15 &  0.8739$\pm$0.05 \\
             & NNDSVD &  0.8748$\pm$0.00 &  0.8686$\pm$0.01 &  0.8764$\pm$0.01 &  0.8740$\pm$0.01 &  0.8713$\pm$0.01 &  0.8693$\pm$0.01 \\
             & FactorAnalysis &  \textbf{0.9997$\pm$0.00} &  \textbf{0.9993$\pm$0.00} &  0.9992$\pm$0.00 &  0.9988$\pm$0.00 &  0.9983$\pm$0.00 &  0.9979$\pm$0.00 \\
             & NMF (CD) &  0.9907$\pm$0.01 &  0.9885$\pm$0.01 &  0.9870$\pm$0.01 &  0.9845$\pm$0.01 &  0.9916$\pm$0.01 &  0.9837$\pm$0.01 \\
             & NMF (MU) &  0.8883$\pm$0.04 &  0.8160$\pm$0.12 &  0.8720$\pm$0.07 &  0.8617$\pm$0.08 &  0.8330$\pm$0.10 &  0.8096$\pm$0.06 \\
             & SNMF &  0.9977$\pm$0.00 &  0.9980$\pm$0.00 &  0.9951$\pm$0.00 &  0.9943$\pm$0.00 &  0.9935$\pm$0.00 &  0.9949$\pm$0.00 \\
             & CoxNMF &  0.9987$\pm$0.00 &  \textbf{0.9993$\pm$0.00} &  \textbf{0.9997$\pm$0.00} &  \textbf{0.9996$\pm$0.00} &  \textbf{0.9996$\pm$0.00} &  \textbf{0.9985$\pm$0.00} \\
    \midrule
    Accuracy & TruncatedSVD &  \textbf{0.8286$\pm$0.04} &  0.8000$\pm$0.03 &  0.8000$\pm$0.03 &  0.8500$\pm$0.04 &  0.8545$\pm$0.02 &  0.8583$\pm$0.02 \\
             & PCA &  0.8249$\pm$0.05 &  \textbf{0.8538$\pm$0.01} &  \textbf{0.8699$\pm$0.01} &  0.8729$\pm$0.01 &  \textbf{0.8742$\pm$0.01} &  0.8263$\pm$0.02 \\
             & SparsePCA &  0.6833$\pm$0.04 &  0.7310$\pm$0.04 &  0.7064$\pm$0.04 &  0.7412$\pm$0.02 &  0.7385$\pm$0.02 &  0.7421$\pm$0.03 \\
             & NNDSVD &  0.7571$\pm$0.04 &  0.8000$\pm$0.03 &  0.8111$\pm$0.03 &  0.8300$\pm$0.04 &  0.8545$\pm$0.02 &  0.8667$\pm$0.03 \\
             & FactorAnalysis &  0.7546$\pm$0.03 &  0.7774$\pm$0.04 &  0.8119$\pm$0.03 &  0.8257$\pm$0.02 &  0.8409$\pm$0.01 &  0.8507$\pm$0.03 \\
             & NMF (CD) &  0.7857$\pm$0.05 &  0.7750$\pm$0.07 &  0.8333$\pm$0.06 &  0.8200$\pm$0.06 &  0.8182$\pm$0.00 &  0.8417$\pm$0.03 \\
             & NMF (MU) &  0.7429$\pm$0.04 &  0.7250$\pm$0.03 &  0.8111$\pm$0.03 &  0.7700$\pm$0.03 &  0.7818$\pm$0.02 &  0.8167$\pm$0.02 \\
             & SNMF &  0.8429$\pm$0.08 &  0.8125$\pm$0.04 &  0.8667$\pm$0.06 &  0.8400$\pm$0.04 &  0.8455$\pm$0.02 &  0.8667$\pm$0.02 \\
             & CoxNMF &  \textbf{0.8286$\pm$0.04} &  0.8500$\pm$0.03 &  0.8667$\pm$0.07 &  \textbf{0.9100$\pm$0.02} &  0.8636$\pm$0.06 &  \textbf{0.8750$\pm$0.04} \\
    \bottomrule
    \end{tabular}
    \caption{Simulation results with multivariate underlying features setup for $K\in\{7,8,9,10,11,12\}$, $\varepsilon = 0$. $\hat{K}$ was searched around $K\pm\{0,1,2\}$ and was determined by highest silhouette score. Experiments repeat 5 times each with random seed $\in\{1,2,3,4,5\}$. Mean values $\pm$ standard deviations were reported, best performed mean values among models were highlighted in bold font.}
    \label{table:main_multivariate_X_noise=0}
\end{table*}

\begin{table*}[t]
    \centering
    \small
    \begin{tabular}{llllllll}
    \toprule
             & $K$  &            7 &            8 &            9 &           10 &           11 &           12 \\
    Metrics & Model &              &              &              &              &              &              \\
    \midrule
    C-Index & TruncatedSVD &  0.8594$\pm$0.07 &  0.8639$\pm$0.13 &  0.8881$\pm$0.11 &  0.8617$\pm$0.17 &  0.8766$\pm$0.10 &  0.9142$\pm$0.05 \\
             & PCA &  0.5685$\pm$0.02 &  0.8695$\pm$0.11 &  0.9865$\pm$0.00 &  0.5842$\pm$0.02 &  0.9871$\pm$0.00 &  0.9153$\pm$0.04 \\
             & SparsePCA &  0.9568$\pm$0.01 &  0.9580$\pm$0.01 &  0.9564$\pm$0.01 &  0.9565$\pm$0.01 &  0.7936$\pm$0.16 &  0.9512$\pm$0.01 \\
             & NNDSVD &  0.5876$\pm$0.03 &  0.5674$\pm$0.03 &  0.5781$\pm$0.02 &  0.5829$\pm$0.03 &  0.5795$\pm$0.03 &  0.6351$\pm$0.04 \\
             & FactorAnalysis &  0.9866$\pm$0.00 &  0.8612$\pm$0.15 &  0.8584$\pm$0.14 &  0.9865$\pm$0.00 &  0.8726$\pm$0.12 &  0.9029$\pm$0.07 \\
             & NMF (CD) &  0.8589$\pm$0.07 &  0.8642$\pm$0.13 &  0.8877$\pm$0.11 &  0.8613$\pm$0.17 &  0.8762$\pm$0.10 &  0.9099$\pm$0.05 \\
             & NMF (MU) &  0.7567$\pm$0.06 &  0.7026$\pm$0.10 &  0.7482$\pm$0.15 &  0.7736$\pm$0.13 &  0.7217$\pm$0.14 &  0.6430$\pm$0.04 \\
             & SNMF &  0.9872$\pm$0.00 &  0.8634$\pm$0.13 &  0.9865$\pm$0.00 &  0.9861$\pm$0.00 &  0.9848$\pm$0.00 &  0.9106$\pm$0.05 \\
             & CoxNMF &  \textbf{0.9990$\pm$0.00} &  \textbf{0.9990$\pm$0.00} &  \textbf{0.9992$\pm$0.00} &  \textbf{0.9993$\pm$0.00} &  \textbf{0.9980$\pm$0.00} &  \textbf{0.9995$\pm$0.00} \\
    \midrule
    Accuracy & TruncatedSVD &  0.6027$\pm$0.05 &  0.7230$\pm$0.09 &  0.7737$\pm$0.02 &  0.7895$\pm$0.02 &  0.7820$\pm$0.06 &  0.7890$\pm$0.05 \\
             & PCA &  0.7113$\pm$0.02 &  0.7374$\pm$0.06 &  0.8528$\pm$0.01 &  0.7951$\pm$0.01 &  \textbf{0.8793$\pm$0.01} &  0.8248$\pm$0.01 \\
             & SparsePCA &  0.6546$\pm$0.04 &  0.8069$\pm$0.03 &  0.7724$\pm$0.05 &  0.7185$\pm$0.05 &  0.7642$\pm$0.05 &  0.7849$\pm$0.03 \\
             & NNDSVD &  0.6924$\pm$0.04 &  0.7119$\pm$0.03 &  0.7424$\pm$0.03 &  0.7580$\pm$0.04 &  0.7724$\pm$0.01 &  0.7889$\pm$0.02 \\
             & FactorAnalysis &  0.7677$\pm$0.03 &  0.7398$\pm$0.06 &  0.7769$\pm$0.04 &  0.8481$\pm$0.03 &  0.8069$\pm$0.05 &  0.8201$\pm$0.04 \\
             & NMF (CD) &  0.6350$\pm$0.04 &  0.7396$\pm$0.07 &  0.7983$\pm$0.04 &  0.7754$\pm$0.02 &  0.7869$\pm$0.03 &  0.8076$\pm$0.04 \\
             & NMF (MU) &  0.6761$\pm$0.00 &  0.7305$\pm$0.05 &  0.7502$\pm$0.03 &  0.8114$\pm$0.03 &  0.7928$\pm$0.03 &  0.8214$\pm$0.02 \\
             & SNMF &  0.8153$\pm$0.04 &  0.7288$\pm$0.10 &  0.8459$\pm$0.05 &  0.8320$\pm$0.02 &  0.8645$\pm$0.02 &  0.8148$\pm$0.04 \\
             & CoxNMF &  \textbf{0.8474$\pm$0.00} &  \textbf{0.8385$\pm$0.07} &  \textbf{0.8583$\pm$0.04} &  \textbf{0.8658$\pm$0.03} &  0.8501$\pm$0.06 &  \textbf{0.8908$\pm$0.02} \\
    \bottomrule
    \end{tabular}
    \caption{Simulation results with multivariate underlying features setup for $K\in\{7,8,9,10,11,12\}$, $\varepsilon = 0.05$. $\hat{K}$ was searched around $K\pm\{0,1,2\}$ and was determined by highest silhouette score. Experiments repeat 5 times each with random seed $\in\{1,2,3,4,5\}$. Mean values $\pm$ standard deviations were reported, best performed mean values among models were highlighted in bold font.}
    \label{table:main_multivariate_X_noise=0.05}
\end{table*}

\begin{table*}[t]
    \centering
    \small
    \begin{tabular}{llllllll}
    \toprule
             & $K$  &            7 &            8 &            9 &           10 &           11 &           12 \\
    Metrics & Model &              &              &              &              &              &              \\
    \midrule
    C-Index & TruncatedSVD &  0.7684$\pm$0.09 &  0.7982$\pm$0.13 &  0.8320$\pm$0.13 &  0.8117$\pm$0.14 &  0.8007$\pm$0.14 &  0.8381$\pm$0.07 \\
             & PCA &  0.7723$\pm$0.09 &  0.8023$\pm$0.12 &  0.5861$\pm$0.04 &  0.9683$\pm$0.00 &  0.7903$\pm$0.14 &  0.8499$\pm$0.06 \\
             & SparsePCA &  0.9458$\pm$0.01 &  0.9468$\pm$0.01 &  0.7893$\pm$0.15 &  0.7954$\pm$0.12 &  0.9473$\pm$0.01 &  0.9416$\pm$0.01 \\
             & NNDSVD &  0.5728$\pm$0.04 &  0.5654$\pm$0.03 &  0.5798$\pm$0.02 &  0.5826$\pm$0.03 &  0.5822$\pm$0.03 &  0.6335$\pm$0.04 \\
             & FactorAnalysis &  0.7893$\pm$0.08 &  0.8059$\pm$0.13 &  0.8116$\pm$0.16 &  0.8163$\pm$0.14 &  0.7973$\pm$0.14 &  0.8440$\pm$0.07 \\
             & NMF (CD) &  0.7669$\pm$0.09 &  0.7979$\pm$0.13 &  0.8318$\pm$0.13 &  0.8125$\pm$0.13 &  0.8019$\pm$0.14 &  0.8309$\pm$0.08 \\
             & NMF (MU) &  0.6994$\pm$0.07 &  0.6947$\pm$0.08 &  0.7326$\pm$0.15 &  0.7294$\pm$0.10 &  0.7073$\pm$0.14 &  0.6472$\pm$0.05 \\
             & SNMF &  0.9695$\pm$0.00 &  0.7982$\pm$0.13 &  0.8305$\pm$0.13 &  0.9676$\pm$0.00 &  0.8013$\pm$0.14 &  0.8293$\pm$0.08 \\
             & CoxNMF &  \textbf{0.9975$\pm$0.00} &  \textbf{0.9982$\pm$0.00} &  \textbf{0.9959$\pm$0.00} &  \textbf{0.9983$\pm$0.00} &  \textbf{0.9993$\pm$0.00} &  \textbf{0.9968$\pm$0.00} \\
    \midrule
    Accuracy & TruncatedSVD &  0.5951$\pm$0.06 &  0.7184$\pm$0.10 &  0.7681$\pm$0.03 &  0.7694$\pm$0.06 &  0.7666$\pm$0.07 &  0.7759$\pm$0.04 \\
             & PCA &  0.7031$\pm$0.03 &  0.7390$\pm$0.07 &  0.7703$\pm$0.01 &  \textbf{0.8688$\pm$0.01} &  0.8215$\pm$0.03 &  0.8209$\pm$0.01 \\
             & SparsePCA &  0.6734$\pm$0.04 &  0.7831$\pm$0.06 &  0.7116$\pm$0.05 &  0.7141$\pm$0.06 &  0.7742$\pm$0.07 &  0.8161$\pm$0.03 \\
             & NNDSVD &  0.6517$\pm$0.02 &  0.7161$\pm$0.03 &  0.7404$\pm$0.03 &  0.7475$\pm$0.03 &  0.7730$\pm$0.01 &  0.7934$\pm$0.02 \\
             & FactorAnalysis &  0.6649$\pm$0.02 &  0.7532$\pm$0.05 &  0.7930$\pm$0.03 &  0.8125$\pm$0.01 &  0.8197$\pm$0.05 &  0.8164$\pm$0.03 \\
             & NMF (CD) &  0.6333$\pm$0.03 &  0.7213$\pm$0.06 &  0.8141$\pm$0.05 &  0.7898$\pm$0.06 &  0.7919$\pm$0.02 &  0.8097$\pm$0.04 \\
             & NMF (MU) &  0.6939$\pm$0.05 &  0.7551$\pm$0.04 &  0.7431$\pm$0.01 &  0.8241$\pm$0.03 &  0.8169$\pm$0.04 &  \textbf{0.8248$\pm$0.02} \\
             & SNMF &  0.8211$\pm$0.03 &  0.7426$\pm$0.06 &  0.7783$\pm$0.02 &  0.8371$\pm$0.03 &  0.7931$\pm$0.02 &  0.7907$\pm$0.02 \\
             & CoxNMF &  \textbf{0.8377$\pm$0.00} &  \textbf{0.8214$\pm$0.03} &  \textbf{0.8203$\pm$0.03} &  0.8625$\pm$0.03 &  \textbf{0.8780$\pm$0.02} &  0.8116$\pm$0.03 \\
    \bottomrule
    \end{tabular}
    \caption{Simulation results with multivariate underlying features setup for $K\in\{7,8,9,10,11,12\}$, $\varepsilon = 0.10$. $\hat{K}$ was searched around $K\pm\{0,1,2\}$ and was determined by highest silhouette score. Experiments repeat 5 times each with random seed $\in\{1,2,3,4,5\}$. Mean values $\pm$ standard deviations were reported, best performed mean values among models were highlighted in bold font.}
    \label{table:main_multivariate_X_noise=0.10}
\end{table*}

\begin{sidewaystable*}[t]
    \resizebox{\textwidth}{!}{%
    \begin{minipage}{2.45\textwidth}
\begin{tabularx}{\textwidth}{llllllllllllllllllllllllllll}
\toprule
   &        & \multicolumn{8}{c}{$\varepsilon=0.00$} & & \multicolumn{8}{c}{$\varepsilon=0.05$} & & \multicolumn{8}{c}{$\varepsilon=0.10$} \\
   &        &        $\hat{K}$ & Relative error &       C-Index &     Accuracy &     F-1 score &    Precision &       Recall & \multicolumn{2}{l}{Runtime (s)} &        $\hat{K}$ & Relative error &       C-Index &     Accuracy &     F-1 score &    Precision &       Recall & \multicolumn{2}{l}{Runtime (s)} &        $\hat{K}$ & Relative error &       C-Index &     Accuracy &     F-1 score &    Precision &       Recall & \multicolumn{1}{l}{Runtime (s)} \\
$K$ & Model &              &                &              &              &              &              &              &                 &   &              &                &              &              &              &              &              &                 &   &              &                &              &              &              &              &              &                \\
\midrule
7  & TruncatedSVD &            \textbf{7} &    \textbf{0.0000$\pm$0.00} &  \textbf{0.9999$\pm$0.00} &  0.9143$\pm$0.08 &  0.8250$\pm$0.16 &  0.8250$\pm$0.16 &  0.8250$\pm$0.16 &     0.4025$\pm$0.04 &   &            6 &    0.0747$\pm$0.00 &  0.8586$\pm$0.07 &  0.7183$\pm$0.08 &  0.5347$\pm$0.12 &  0.5352$\pm$0.10 &  0.5498$\pm$0.15 &     \textbf{0.0873$\pm$0.00} &   &            6 &    0.1180$\pm$0.01 &  0.7746$\pm$0.10 &  0.7140$\pm$0.07 &  0.5301$\pm$0.12 &  0.5313$\pm$0.10 &  0.5457$\pm$0.14 &     0.0706$\pm$0.02 \\
\bottomrule
\end{tabularx}
\end{minipage}
}
    \caption{Simulation results with univariate underlying features setup among all combinations of $K\in\{7,8,9,10,11,12\}$ and $\varepsilon\in\{0,0.05,0.10\}$. Experiments repeat 5 times each with random seed $\in\{1,2,3,4,5\}$. Mean values $\pm$ standard deviations were reported, best performed mean values among models were highlighted in bold font.}
    \label{table:supplement_univariate}
\end{sidewaystable*}

\begin{sidewaystable*}[t]
    \resizebox{\textwidth}{!}{%
    \begin{minipage}{2.45\textwidth}
    \begin{tabularx}{\textwidth}{llllllllllllllllllllllllllll}
\toprule
   &        & \multicolumn{8}{c}{$\varepsilon=0.00$} & & \multicolumn{8}{c}{$\varepsilon=0.05$} & & \multicolumn{8}{c}{$\varepsilon=0.10$} \\
   &        &        $\hat{K}$ & Relative error &       C-Index &     Accuracy &     F-1 score &    Precision &       Recall & \multicolumn{2}{l}{Runtime (s)} &        $\hat{K}$ & Relative error &       C-Index &     Accuracy &     F-1 score &    Precision &       Recall & \multicolumn{2}{l}{Runtime (s)} &        $\hat{K}$ & Relative error &       C-Index &     Accuracy &     F-1 score &    Precision &       Recall & \multicolumn{1}{l}{Runtime (s)} \\
$K$ & Model &              &                &              &              &              &              &              &                 &   &              &                &              &              &              &              &              &                 &   &              &                &              &              &              &              &              &                \\
\midrule
7  & TruncatedSVD &            \textbf{7} &    \textbf{0.0000$\pm$0.00} &  \textbf{0.9997$\pm$0.00} &  0.8286$\pm$0.04 &  0.7078$\pm$0.07 &  0.7542$\pm$0.08 &  0.6848$\pm$0.06 &     0.3964$\pm$0.01 &   &            6 &    0.0755$\pm$0.00 &  0.8594$\pm$0.07 &  0.6027$\pm$0.05 &  0.4159$\pm$0.06 &  0.4181$\pm$0.06 &  0.4141$\pm$0.06 &     \textbf{0.0881$\pm$0.00} &   &            6 &    0.1161$\pm$0.00 &  0.7684$\pm$0.09 &  0.5951$\pm$0.06 &  0.4195$\pm$0.07 &  0.4241$\pm$0.06 &  0.4158$\pm$0.07 &    0.0575$\pm$0.02 \\
\bottomrule
\end{tabularx}
\end{minipage}
}
    \caption{Simulation results with multivariate underlying features setup among all combinations of $K\in\{7,8,9,10,11,12\}$ and $\varepsilon\in\{0,0.05,0.10\}$. Experiments repeat 5 times each with random seed $\in\{1,2,3,4,5\}$. Mean values $\pm$ standard deviations were reported, best performed mean values among models were highlighted in bold font.}
    \label{table:supplement_multivariate}
\end{sidewaystable*}

\begin{table*}[ht!]
    \centering
    \small
    \begin{tabular}{ c | c p{40mm} | c c c c }
    \toprule
     & \multicolumn{6}{c}{Gene cluster $C2$ associated with better survival prognosis}\\
    \midrule
    Rank & Term & Description & Input genes & Genes in GO term & Genes overlapped & $P$-value\\
    \midrule
    1 & GO:0016071 & mRNA metabolic process. & 1702 & 901 & 277 & $1.244\times 10^{-90}$ \\
    2 & GO:0072599 & Establishment of protein localization to endoplasmic reticulum (ER). & 1702 & 124 & 89 & $1.982\times 10^{-68}$ \\
    3 & GO:0045047 & Protein targeting to ER. & 1702 & 120 & 85 & $1.353\times 10^{-64}$ \\
    \midrule
    & \multicolumn{6}{c}{Gene cluster $C14$ associated with worse survival prognosis}\\
    \midrule
    Rank & Term & Description & Input genes & Genes in GO term & Genes overlapped & $P$-value\\
    \midrule
    1 & GO:0090630 & Activation of GTPase activity. & 337 & 101 & 27 & $1.698\times 10^{-25}$\\
    2 & GO:1902017 & Regulation of cilium assembly. & 337 & 87 & 24 & $3.770\times 10^{-23}$\\
    3 & GO:1903508 & Positive regulation of nucleic acid-templated transcription. & 337 & 1736 & 78 & $8.182\times 10^{-17}$\\
    \bottomrule
    \end{tabular}
    \caption{Gene ontology (GO) enrichment analysis results for Kidney Renal Clear Cell Carcinoma (KIRC). Top 3 ranked GO terms were reported according to the $P$-values, which were associated with better/worse survival prognosis.}
    \label{table:KIRC}
\end{table*}

\begin{table*}[t]
    \centering
    \small
    \begin{tabular}{ c | c p{40mm} | c c c c }
    \toprule
     & \multicolumn{6}{c}{Gene cluster $C8$ associated with better survival prognosis}\\
    \midrule
    Rank & Term & Description & Input genes & Genes in GO term & Genes overlapped & $P$-value\\
    \midrule
    1 & GO:0006886 & Intracellular protein transport. & 748 & 1536 & 135 & $1.254\times 10^{-22}$ \\
    2 & GO:0090630 & Activation of GTPase activity. & 748 & 101 & 28 & $1.489\times 10^{-17}$ \\
    3 & GO:0097503 & Sialylation. & 748 & 20 & 14 & $1.849\times 10^{-16}$ \\
    \midrule
    & \multicolumn{6}{c}{Gene cluster $C3$ associated with worse survival prognosis}\\
    \midrule
    Rank & Term & Description & Input genes & Genes in GO term & Genes overlapped & $P$-value\\
    \midrule
    1 & GO:0019637 & Organophosphate metabolic process. & 1160 & 1203 & 238 & $6.104\times 10^{-71}$\\
    2 & GO:0046488 & Phosphatidylinositol metabolic process. & 1160 & 221 & 80 & $1.034\times 10^{-43}$\\
    3 & GO:0006650 & Glycerophospholipid metabolic process. & 1160 & 366 & 102 & $1.337\times 10^{-43}$\\
    \bottomrule
    \end{tabular}
    \caption{Gene ontology (GO) enrichment analysis results for Lung Adenocarcinoma (LUAD). Top 3 ranked GO terms were reported according to the $P$-values, which were associated with better/worse survival prognosis.}
    
    \label{table:LUAD}
\end{table*}

\end{document}